\documentclass[conference]{IEEEtran}
\let\labelindent\relax %

\renewcommand{\paragraph}[1]{\noindent\textbf{#1.}\hspace{1ex}}

\usepackage{paralist}
\usepackage{fancyhdr}
\usepackage{xspace}
\usepackage{hyperref}
\usepackage{cleveref}
\usepackage{makecell}
\usepackage{booktabs}
\usepackage{multirow}
\let\labelindent\relax %
\usepackage{enumitem}
\usepackage{calc}
\usepackage{array}
\usepackage{listings}
\usepackage{tabularx}
\usepackage{algorithm}
\usepackage{algorithmic}
\usepackage{subfigure}
\usepackage{etoolbox}
\usepackage{balance}
\newtoggle{draftmode}
\settoggle{draftmode}{false}

\newtoggle{camera}
\settoggle{camera}{true}

\usepackage{tikz}

\newcommand{\coolname}{\textsc{DARWIN}\xspace}

\newcommand{\changed}[1]{\iftoggle{draftmode}{\textcolor{blue}{#1}}{#1}}

\definecolor{mGreen}{rgb}{0,0.6,0}
\definecolor{mGray}{rgb}{0.5,0.5,0.5}
\definecolor{mPurple}{rgb}{0.58,0,0.82}
\definecolor{backgroundColour}{rgb}{0.95,0.95,0.92}

\lstdefinestyle{CStyle}{
    backgroundcolor=\color{backgroundColour},   
    commentstyle=\color{mGreen},
    keywordstyle=\color{magenta},
    numberstyle=\tiny\color{mGray},
    stringstyle=\color{mPurple},
    basicstyle=\footnotesize,
    breakatwhitespace=false,         
    breaklines=true,                 
    captionpos=b,                    
    keepspaces=true,                 
    numbers=left,                    
    numbersep=5pt,                  
    showspaces=false,                
    showstringspaces=false,
    showtabs=false,                  
    tabsize=2,
    language=C
}

\crefname{lstlisting}{listing}{listings}
\Crefname{lstlisting}{Listing}{Listings}

\begin{document}

\date{}

\title{\coolname: Survival of the Fittest Fuzzing Mutators}

\iftoggle{camera}%
{%
\author{\IEEEauthorblockN{Patrick Jauernig\IEEEauthorrefmark{1}, Domagoj Jakobovic\IEEEauthorrefmark{3}, Stjepan Picek\IEEEauthorrefmark{4}, Emmanuel Stapf\IEEEauthorrefmark{1} and Ahmad-Reza Sadeghi\IEEEauthorrefmark{2}}
    \IEEEauthorblockA{\IEEEauthorrefmark{1}Technical University of Darmstadt, Germany, \{patrick.jauernig, emmanuel.stapf\}@sanctuary.dev}
    \IEEEauthorblockA{\IEEEauthorrefmark{2}Technical University of Darmstadt, Germany, ahmad.sadeghi@trust.tu-darmstadt.de}
	\IEEEauthorblockA{\IEEEauthorrefmark{3}University of Zagreb, Croatia, domagoj.jakobovic@fer.hr}
	\IEEEauthorblockA{\IEEEauthorrefmark{4}Radboud University and TU Delft, The Netherlands, picek.stjepan@gmail.com}%
}
} {\author{}}

\maketitle

\begin{abstract}
Fuzzing is an automated software testing technique broadly adopted by the industry.
A popular variant is mutation-based fuzzing, which discovers a large number of bugs in practice.
While the research community has studied mutation-based fuzzing for years now, the algorithms' interactions within the fuzzer are highly complex and can, together with the randomness in every instance of a fuzzer, lead to unpredictable effects.
Most efforts to improve this fragile interaction focused on optimizing seed scheduling.
However, real-world results like Google's FuzzBench highlight that these approaches do not consistently show improvements in practice.
Another approach to improve the fuzzing process algorithmically is optimizing mutation scheduling.
Unfortunately, existing mutation scheduling approaches also failed to convince because of missing real-world improvements or too many user-controlled parameters whose configuration requires expert knowledge about the target program.
This leaves the challenging problem of cleverly processing test cases and achieving a measurable improvement unsolved.\\
We present \coolname, a novel mutation scheduler and the first to show fuzzing improvements in a realistic scenario without the need to introduce additional user-configurable parameters, opening this approach to the broad fuzzing community. \coolname uses an Evolution Strategy to systematically optimize and adapt the probability distribution of the mutation operators during fuzzing. 
We implemented a prototype based on the popular general-purpose fuzzer AFL. 
\coolname significantly outperforms the state-of-the-art mutation scheduler and the AFL baseline in our own coverage experiment, in FuzzBench, and by finding 15 out of 21 bugs the fastest in the MAGMA benchmark. 
Finally, \coolname found 20 unique bugs (including one novel bug), 66\% more than AFL, in widely-used real-world applications.
\end{abstract}
\section{Introduction}
\label{sec:introduction}

Vulnerabilities caused by programming errors are still a major threat to today's programs~\cite{owasp}.
An important class of programming errors is memory corruption vulnerabilities, where unexpected, malformed inputs can lead to uncontrolled behavior in the program, which can often be abused by attackers.
A modern, cost-efficient strategy to uncover these programming errors is automated software testing using fuzz testing (commonly known as \textit{fuzzing}).
Fuzzing automatically generates inputs from testcases and feeds them to the program under test while monitoring the program.
If a programming error has been reached, the fuzzer notices that the program hangs or crashes.
Optionally, the observed control-flow changes can serve as feedback for the next iteration, i.e., whether a new path in the control flow (known as \textit{coverage}) has been taken due to the generated input.
In recent years, fuzzers emerged as an important topic in academic as well as industrial research and are nowadays widely used for finding bugs in commercial software~\cite{ms-onefuzz,google-ossfuzz}.
Projects like Google OSSFuzz~\cite{google-ossfuzz} helped to significantly increase the adoption rate by offering free computation for fuzzing while still allowing security researchers, who provide the fuzzers, to keep the bug bounty for discovered vulnerabilities.

While fuzzers are responsible for discovering tremendous amounts of bugs, even in operating system kernels~\cite{syzkaller}, they are still extensively researched, e.g., in the areas of making targets available to fuzz testing~\cite{firmwarefuzzing,firmwarefuzzing2}, improving fuzzers using new algorithms~\cite{aflgo,hawkeye,steelix,mopt,fairfuzz,rajpal2017not}, and leveraging new hardware features for performance or coverage improvements~\cite{kAFL,chen2019ptrix}.

This paper focuses on the subject of algorithmic improvements for mutational fuzzers, which leverage an existing set of testcases (referred to as \textit{corpus}) to constantly generate new variants of these testcases by applying \textit{mutation operators} inspired by genetic mutations.
Most notably, a significant number of works focused on the effects of algorithmically sampling a subset of optimal seeds from the corpus.
The goals of these works range from removing redundancy to creating a minimal coverage-preserving corpus with small files~\cite{ecofuzz,moonshine}, efficiently reaching specific locations in the control-flow graph~\cite{aflgo,hawkeye,fuzzguard}, or improving coverage in general~\cite{fasterfuzzing}.
While these approaches are designed to select from a large number of possible testcases, in reality, testcases suitable for fuzzing are often rare~\cite{afl}.

Aside from seed-selection algorithms, other approaches have been proposed~\cite{fairfuzz,steelix,rajpal2017not} that approximate which byte positions in the testcase give the best results when being mutated, but not which mutation operators to apply.
Yet, this problem is highly challenging, as it is required to be shown whether 1) mutation selection is actually target-dependent, 2) the selection distribution is static or dynamic, 3) introducing an optimization algorithm reduces execution speed s.t. its better mutation selection is outweighed.

The first approach to optimize the actual selection of mutations (\textit{mutation scheduling}) has been MO\small{PT}\normalsize~\cite{mopt}.
MO{\small PT} \normalsize proposes a variant of the Particle Swarm Optimization algorithm (PSO) to learn a globally optimal mutation probability distribution.
However, MO{\small PT}\normalsize{}'s PSO algorithm has both local and global best probability distributions, making finding the best solution, and therefore the algorithm itself, complex and more expensive to use during fuzzing.
Similar to other algorithmic improvements to fuzzing, finding a practical trade-off between complexity and algorithmic improvements is challenging. All additional algorithms have direct implications on execution speed, and hence, reduce coverage over time.
Further, MO{\small PT} \normalsize introduces various user-configurable parameters that steer the optimization process directly, so the user needs to solve another complex problem instead to avoid non-optimal scheduling.
For a reasonable choice of parameters, the user either needs expert knowledge of the target application or a preliminary fuzzing campaign.
Finally, MO{\small PT} \normalsize fails to outperform AFL, which is built on, in the popular FuzzBench fuzzer benchmark by Google~\cite{fuzzbench-report}.
This makes designing and building a practical mutation scheduler a challenging open problem.

\textbf{This work.} This paper focuses on one aspect of the fuzzing process: finding (approx.) optimal mutation scheduling strategies to improve fuzzing algorithmically.
Here, the challenging goal is to infer which mutation operator is the optimal choice for the next fuzzing iteration.
We address this problem with \coolname, a novel mutation-scheduling algorithm to improve the general performance of mutational fuzzers.
\coolname leverages an Evolution Strategy (ES), in a setting similar to reinforcement learning, to approximate ideal probability distribution for the mutation operator selection to avoid wasting fuzzing iterations on suboptimal mutators.
The resulting probability distribution is not statically set but learned during the fuzzing process and dynamically adapted to the target program.
\coolname outperforms related work significantly, not only in coverage but also in the time to find bugs, without the user having to adjust any target-specific parameters, which allows non-expert users to leverage mutation scheduling.\\

\textbf{Challenges.} Although we focus only on a specific phase of the fuzzing process, namely the mutation selection in the havoc phase, the problem of finding an optimal probability distribution for mutation selection is highly challenging: numerous mutation operators can be used, and their efficiency varies depending on the target program, the current input, and the state inherently implied by the current input.
Furthermore, the efficiency can vary depending on the non-deterministic nature of each fuzzing run and the interplay between fuzzing stages.
Therefore, it is impossible to examine all possible options exhaustively in the general case.\\

\textbf{Contributions.} 
Our \coolname mutation scheduler and its implementation based on AFL tackle all these challenges.
To summarize, our main contributions include:

\begin{compactitem}
    \item We present a novel mutation scheduling approach, \coolname; the first mutation scheduler that leverages a variant of Evolution Strategy to optimize the probability distribution of mutation operators. \coolname dramatically improves the efficiency of mutation selection while keeping the execution speed constant. \coolname can be applied to any feedback-guided mutation-based fuzzer.
    \item We implemented a prototype of \coolname by extending AFL with our mutation scheduling algorithm. By modifying only three code lines in AFL to integrate our \coolname mutation scheduler, we show that \coolname's design is easily adoptable by existing fuzzers. We further highlight this by also integrating \coolname in EcoFuzz\cite{ecofuzz}. What is more, we do not introduce any additional user-configurable parameters to avoid creating adoption barriers.
    \item We thoroughly evaluate \coolname against AFL as a baseline and the most recent related work in this area, MO\small{PT}\normalsize{}. Our prototype significantly outperforms both fuzzers, MO\small{PT}\normalsize{} and AFL, in terms of code coverage reached in the well-fuzzed GNU binutils suite. Next, \coolname is the first mutation scheduler to outperform its base fuzzer in Google's Fuzzbench. Further, we evaluate \coolname on MAGMA, where we show that \coolname triggers 15 out of the 21 bugs found the fastest. Finally, \coolname finds 20 unique bugs (including one previously unreported bug), 66\% more than AFL, across various real-world targets.
    \item We thoroughly analyze the root causes for \coolname's efficiency by first comparing \coolname to a static pre-optimized mutation probability distribution, and further, studying the mutation probability distribution over time, and introducing a metric to measure a fuzzer's effectiveness in scheduling mutations. We show that \coolname needs fewer mutations than AFL to reach a coverage point while achieving a higher execution speed than the state-of-the-art MO\small{PT}\normalsize{} fuzzer.
\end{compactitem}

To foster future research in this area, we open-source our fuzzer at \url{https://github.com/TUDA-SSL/DARWIN}.

\section{Background}
This section presents the necessary background information to understand the general concept of fuzzers, the workflow of mutation-based fuzzers, and metaheuristic optimization.

\subsection{Fuzzing}

On a high level, fuzzers can be divided into mutational, i.e., mutating testcases, and generational, i.e., deriving structured inputs, fuzzers.
Mutational fuzzing requires a set (\textit{corpus}) of program inputs (\textit{seeds}), which can, e.g., be obtained from testcases or real inputs.
These seeds are then mutated using operations known from genetics, like inserting random errors (bit flips), changing values to corner cases, or combining two inputs to create a new input.
As this way of input generation does not follow any constraints on the input, the generated inputs are more unlikely to pass, e.g., initial parser checks or checksums~\cite{TaintScope}.
The process of mutation can be influenced in two ways: 1) the location in the input that gets mutated and 2) the mutation that is applied, whereby the selection can either be made randomly or guided by a heuristic.
Such a heuristic can be, e.g., success measured in an increase of coverage or a certain state that should be reached (where the target has been tainted to find a clear path to that state).
For example, the popular AFL fuzzer uses the coverage metric of basic-block transitions as a heuristic~\cite{afl}.

\subsection{Fuzzing Loop of Mutational Fuzzers}

For mutational fuzzers, the so-called fuzzing loop, which is the place in the code where the loaded seeds are mutated before being used as inputs for the program under test, usually can be divided into three stages, the deterministic, havoc, and splicing stage~\cite{kAFL,aflgo,hawkeye,mopt,redqueen}. While some aspects are AFL-specific, most concepts presented are implemented in a similar way for other fuzzers. 

\textbf{Deterministic stage.} In the first stage, the deterministic stage, a small set of mutations is applied to seeds in a predefined order to create inputs for the target program, whereby the seeds are drawn from a queue of initial seeds provided by the user. AFL uses code coverage as a heuristic to decide whether a mutated seed has been \textit{successful}. If a seed increases the code coverage, it is stored in the \textit{fuzzing queue}. By reusing successful seeds in the later iterations, the overall fuzzing performance is improved.
Measuring the code coverage is achieved by instrumenting the binary of the program under test such that the program is intercepted on every branch hit.
When an input leads to a crash of the program under test, the user is notified since this indicates a bug in the program. 
The first stage of the fuzzing loop with its deterministic mutation scheme is slow and tends to contribute less to the overall coverage~\cite{mopt}. Thus, AFL allows disabling the deterministic stage entirely, which is especially beneficial for short fuzzing runs~\cite{mopt} or to reduce noise in performance measurements of the following stages.

\textbf{Havoc stage.} In the second stage of the fuzzing loop, the non-deterministic havoc stage, randomly chosen mutations are selected from a list of mutational operators~\cite{libfuzzer,vuzzer,afl,angora}. In~\Cref{tab:mut_operators}, Appendix~\ref{app:afl_mutations}, we list the mutations used in AFL's havoc stage. The selected mutations are applied to the inputs received from the deterministic stage or to the mutated seeds from the fuzzing queue. When the generated program inputs achieve new coverage, they are again saved in the fuzzing queue. The fuzzing loop then returns to the deterministic stage and selects the next element from the fuzzing queue for the next iteration. 
The havoc stage is the most generic stage and widely adopted by AFL-based and other mutational fuzzers~\cite{vuzzer,angora,afl,afl++}, which is also why our novel mutation scheduler \coolname targets the havoc stage.

\textbf{Splicing stage.} The last stage of the fuzzing loop, the splicing stage, is only activated when none of the inputs in the fuzzing queue led to new coverage in the havoc stage. In the splicing stage, a crossover mutation of two inputs is performed, which is then fed back to the havoc stage, which again applies a random mutation on the input before testing it on the target program.

\subsection{Metaheuristics}
\label{subsec:metaheuristics}

While in the previous section, we mentioned several approaches to fuzzing, we did not discuss how such approaches can actually find good solutions.
This is because there exist no specialized algorithms developed for that particular problem. Instead, we need to rely on more general solving procedures.
Metaheuristics represent an intuitive choice since they encompass problem-independent techniques used in a broad range of applications.
For example, we can consider the problem of finding a suitable \textit{mutation schedule} in the havoc stage as an optimization problem.
Since there is no explicit cost function for this optimization problem, it cannot readily be paired with classical optimization algorithms requiring gradient information.
In that case, metaheuristic algorithms, which do not pose any requirements on the optimization problem, have proven to be the method of choice in many engineering applications. Metaheuristic techniques are commonly used in domains like the automotive industry~\cite{doi:10.1080/0305215X.2019.1651310}, medicine~\cite{Abouhawwash572}, scheduling~\cite{7101236}, adversarial examples~\cite{8601309}, and implementation attacks~\cite{10.1007/978-3-030-40186-3_8}.

Metaheuristics, in their original definition, represent solution finding methods that orchestrate an interaction between local improvement and higher-level strategies to create a process capable of escaping from local optima and performing a robust search in a solution space~\cite{meta}. A common division of metaheuristic optimization algorithms is into single solution-based and population-based metaheuristics~\cite{Talbi}.
Population-based metaheuristics work on a population of solutions (e.g., Evolutionary Algorithms (EA) and swarm algorithms like Particle Swarm Optimization (PSO)).
A population in this context denotes a set of individuals used during an optimization process, whereby an individual is a data structure that corresponds to an element in the search space (a candidate solution).
In contrast, single solution-based metaheuristics manipulate and transform a single solution (or a smaller number of solutions) during the search.

Evolutionary algorithms occupy a prominent place among metaheuristic algorithms, as they have been successfully applied to a large number of difficult optimization problems~\cite{9185861,10.1007/978-3-030-58115-2_4}.
We depict pseudocode for the generic evolutionary algorithm in Algorithm~\ref{alg:ea}, Appendix~\ref{app:ea}. 
In each iteration, the algorithm applies a selection mechanism that emulates natural selection.
Based on their respective quality, usually denoted as \textit{fitness}, better individuals survive, while worse ones are eliminated.
The population then undergoes variation, creating new genetic material as new individuals in the population.
Finally, all the individuals are reevaluated, and the process is repeated until a specific termination criterion is met.
Since no knowledge is presumed about the nature of the solutions in the current population, the termination is usually based on the number of iterations, allotted time, or finding a solution of acceptable quality.

\begin{figure*}[t]
 	\centering
 	\includegraphics[width=0.6\textwidth]{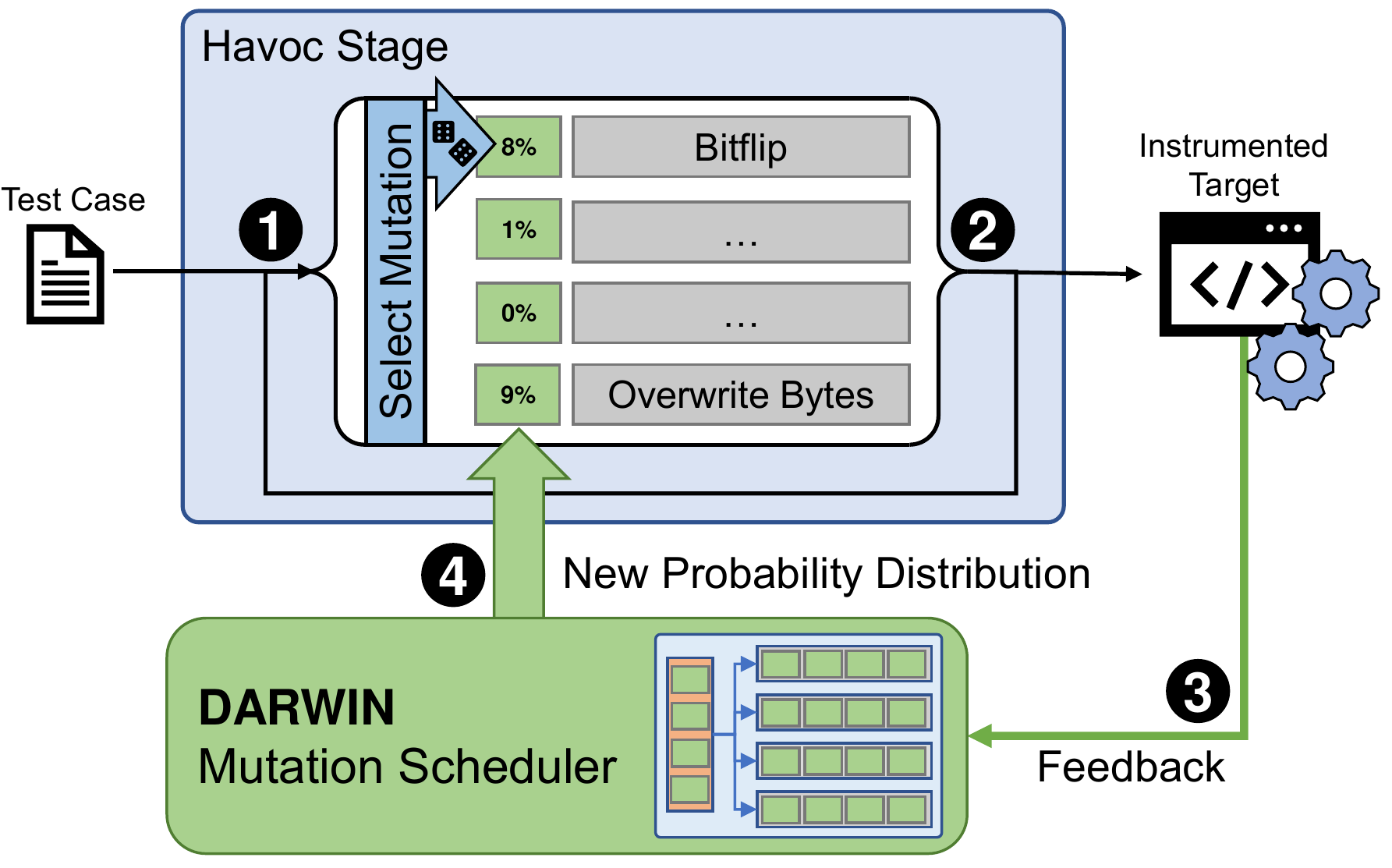}
 	\caption{High-level overview showing how \coolname iteratively optimizes the probability distribution for mutation selection and how the selected mutations are applied to the testcases.}
 	\label{fig:overview}
\end{figure*}

Metaheuristic optimization algorithms balance diversification and intensification properties; diversification enables the discovery of promising areas in the search space and escaping from local optima.
Intensification aims to exploit a promising area by concentrating on the current best solution and finding better neighboring solutions.
The interplay of these properties determines the effectiveness of metaheuristic methods when applied to a specific optimization problem.
\section{Challenges}
\label{sec:challenges}

Designing a mutation scheduling algorithm comes with a number of challenges, as mutation scheduling is a fragile part in the fuzzing process. These challenges are:

\begin{enumerate}[itemsep=0mm, label=\textbf{C.\arabic*:}, ref={C.\arabic*}, leftmargin=0ex, labelsep=\widthof{~}, itemindent=\widthof{C.2:~\hspace{.125ex}}]

\item \label{itm:optimal} \textbf{Optimal Mutation Selection.}
Finding an optimal probability distribution for mutation selection is challenging, as the optimal distribution might change per target.
Further, the probability distribution might depend on the state implied by a part of the input (that is not mutated). 
Hence, a mutation scheduler needs to show that this mutation selection indeed needs to adapt dynamically and, if so, show that iterative adaption outperforms random selection.

\item \label{itm:algo} \textbf{Integrating an Optimization Algorithm.}
Properly selecting a candidate algorithm for mutation scheduling is itself highly challenging.
However, integrating this algorithm into the existing fuzzing process requires a 1) carefully designed representation not only of the problem but also the solution to avoid spending too much computation on encoding, 2) finding a parameter fit for the respective algorithm that fine-tunes exploration versus intensification.

\item \label{itm:adoption} \textbf{Easy Adoption and Reproducibility.}
A complex approach with a large number of user-tweakable parameters might achieve outstanding results. However, it will still not be used in practice due to the difficulties in integrating the approach into a fuzzer or because users fail to find good parameter values, and hence, they cannot achieve results similar to the ones reported by the authors.

\item \label{itm:cperf} \textbf{Performance Trade-off.}
Achieving an optimal trade-off for the mutation selection scenario, which is our goal, is complex.
For instance, fuzzing approaches typically tune the trade-off between performance and cleverness in seed selection. Better seeds reach basic blocks guarded by complex constraints, but optimizing seed selection with algorithms takes additional time, and hence, decreases execution speed.

\end{enumerate}

We designed \coolname with these challenges in mind. Next, we explain how we addressed these challenges throughout the design, implementation, and evaluation of \coolname.
\section{\coolname Design}

\coolname is a novel mutation scheduling algorithm using an Evolution Strategy (ES) to find an optimal mutation selection probability distribution to be applied during the havoc stage.
\coolname is not only determining a static probability distribution but keeps on adapting the distribution throughout the fuzzing run based on coverage information.
Our approach, as depicted in~\Cref{fig:overview}, comprises a well-defined optimization module that does not need to expose any parameters to the user of the fuzzer.
In detail, a fuzzer featuring \coolname performs the following steps in the havoc stage (each step is marked in~\Cref{fig:overview}):\\
\begin{compactenum}
    \item At the beginning of the havoc stage, the fuzzer selects an input from the queue and randomly selects the next mutation to apply. Initially, the probability distribution for mutation selection is uniform.
    \item After applying a mutation, the fuzzer decides whether it should keep mutating this input or if the input should be tested on the instrumented application.
    \item After running the instrumented application with the selected input, feedback is reported to assign a success score to the test input and the \coolname Mutation Scheduler. The mutation scheduler learns based on the reported feedback and optimizes the probability distribution using \coolname's Evolution Strategy. 
    \item Finally, the updated probability distribution is applied for the next iteration.\\
\end{compactenum}

In the following, we explain the optimization process of \coolname's Mutation Scheduler in more detail.

\subsection{Metaheuristics and Mutation Scheduling}

In the context of the complete fuzzing pipeline, we concentrate on improving the mutation scheduler, as illustrated in Figure~\ref{fig:overview}.
The problem of finding a suitable mutation schedule is considered here as an optimization problem, where the candidate solution is a vector of relative mutation operator probabilities.
In a classical optimization scenario, a candidate solution is refined through a series of iterations.
In each iteration, the candidate is evaluated, which is usually the most time-consuming part of the optimization.
Only after a number of iterations, when a candidate of acceptable quality is obtained, the solution is applied to the process being optimized.

In the case of fuzzing, however, the optimization is performed concurrently with the process being optimized since each candidate solution is used as it is being evaluated, and the optimization is performed for each target independently.
Because of this, the optimization algorithm should be able to provide a fast convergence, which means as large a performance improvement with as few evaluations as possible.

As mentioned in Section~\ref{subsec:metaheuristics}, metaheuristic techniques balance between diversification and intensification, with conflicting goals to evade local optima and, at the same time, enable convergence to better quality solutions.
Population-based metaheuristics, such as Genetic Algorithm (GA)~\cite{10.5555/522098} or Particle Swarm Optimization (PSO)~\cite{Kennedy2010} are generally focused on diversification and can locate an optimum with a greater probability.
However, as mentioned above, the optimum in the fuzzing process is not fixed, and the algorithm should adapt swiftly to the current target.
Since they need to evaluate a population of candidate solutions in every step, these approaches usually require a large number of evaluations, and consequently, computation time, to reach a solution of acceptable quality.
Those methods may also include computationally intense domain-dependent operators acting on multiple solutions, such as the crossover operator in GAs, which is a process where a new individual is created from two or more parent solutions~\cite{Eiben03}.
Since, in our case, fast convergence and ease of use are the primary goals, population-based metaheuristics do not present an appropriate choice.

Instead of population-based methods, algorithms that operate on a single solution (or a small set of solutions) should prove to be a better option.
It is expected that single solution algorithms will obtain better performance; since they primarily focus on intensification, convergence is usually faster than in the population-based methods~\cite{convergence}.

In optimizing fuzzing mutation probabilities, where each evaluation may take a considerable amount of time, this behavior translates into a far smaller number of evaluations needed to reach an acceptable solution quality.
At the same time, such algorithms still provide a means to escape local optima with solution perturbations and random restarts.
Examples of these algorithms include Simulated Annealing (SA)~\cite{10.5555/59580}, Tabu Search (TS)~\cite{10.5555/549765}, and Evolution Strategy (ES)~\cite{10.1023/A:1015059928466}. %

\subsection{Evolution Strategy as used in \coolname}

\begin{figure} 	
\centering
 	\includegraphics[width=\linewidth]{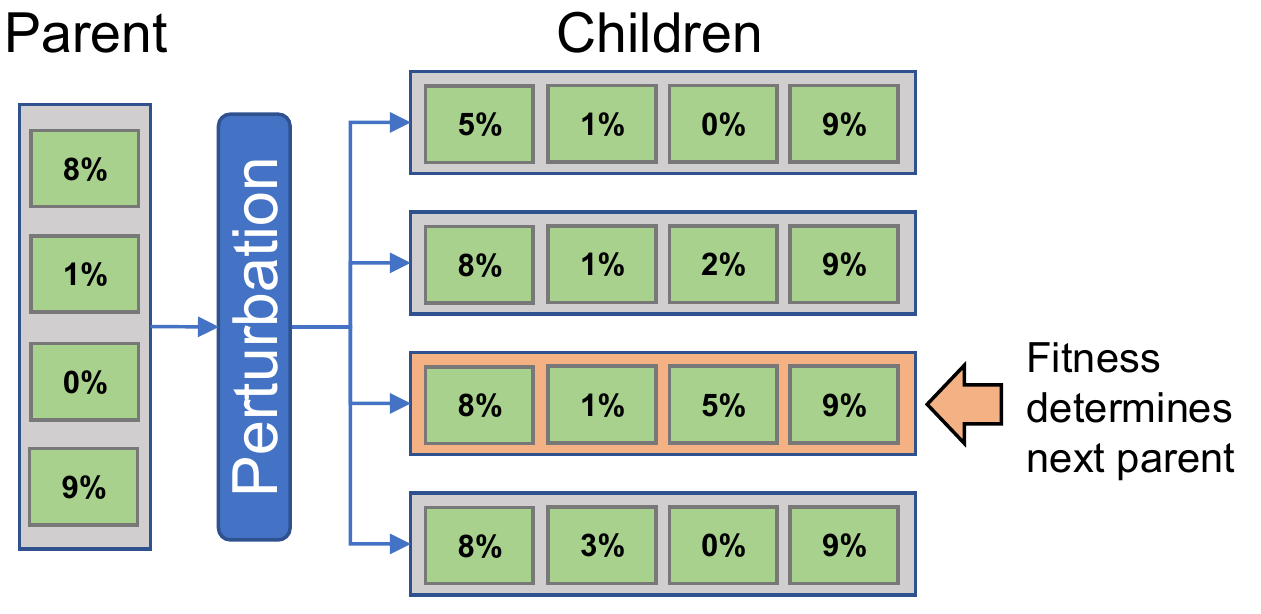}
 	\caption{Example of an ES instantiation with one parent and four children ($\mu=1, \lambda=4$). Based on the fitness function, the parent for the next iteration is determined.}
 	\label{fig:overview_es}
\end{figure}

When considering domain-independent optimization methods, as is the case here, Evolution Strategy has proven to be an efficient and versatile method found in a multitude of applications~\cite{emmerich2018evolution,10.1080/23311916.2014.945820}.
As such, we opted to use ES as the method of choice, both for its simplicity and proven track record as a multi-purpose optimization algorithm~\cite{10.1023/A:1015059928466}. Additionally, ES is well-known to be robust~\cite{1688465}, making it an ideal choice when dealing with difficult optimization problems.

The intensification process in metaheuristics is commonly performed with the use of a mutation operator. 
Mutation operators use only one parent and create one child by applying a randomized change to its genotype (i.e., the encoding of an object)~\cite{Eiben03}. 
However, since we already use the term ``mutation'' for changes in seeds performed by the fuzzer, we will slightly bend the terminology and denote the mutation operator used in ES as the \textit{perturbation} operator. We depict the process in Figure~\ref{fig:overview_es}. %

In its most common form, ES operates on a single solution $\mu$, called the parent.
In each iteration, a randomized perturbation operator is applied on the parent solution producing a number of different modified solutions, commonly called children.
The number of children solutions is denoted with $\lambda$, which is a parameter of the algorithm.
After every child solution is evaluated, the best among all the children solutions and the current parent is chosen as the parent in the next iteration.
This allows \coolname to adjust the mutation schedule dynamically, addressing Challenge~\ref{itm:optimal}.
This type of Evolution Strategy is denoted as $(\mu + \lambda)-ES$. If the parent is disregarded, such selection method is denoted as $(\mu, \lambda)-ES$.
The process is repeated until a designated termination criterion is met, commonly based on elapsed time or a number of evaluations.
We provide the ES pseudocode in Algorithm~\ref{alg:es}.

\begin{algorithm}
\small
\caption{Evolution Strategy\label{alg:es}}
\begin{algorithmic}[1]
  \STATE initialize the parent solution
  \REPEAT
    \STATE create $\lambda$ child solutions using perturbation on the parent
    \STATE select the best solution
    \STATE set the best solution as the parent
  \UNTIL{\textit{TerminationCriterion}}
\end{algorithmic}
\end{algorithm}

When using a single starting parent solution, the algorithm will mainly concentrate on its relative vicinity in the search space. While it is possible for the perturbation operator to move the search to a more distant area, this occurs with a lower probability.
To allow the fuzzer to discover more promising areas in the search space (e.g., more efficient mutation operator combinations), \coolname uses an extended form of the algorithm that starts not with one but several different starting parent solutions.
In this case, the search is conducted in parallel, independently for each parent, addressing Challenge~\ref{itm:cperf}. %
The number of parent solutions in this algorithm variant is denoted with $\mu$.
The modified algorithm can be represented with the following pseudocode:

\begin{algorithm}
\small
\caption{Multi-parent Evolution Strategy\label{alg:mpes}}
\begin{algorithmic}[1]
  \STATE initialize $\mu$ parent solutions
  \REPEAT
  \FORALL{parent solutions}
    \STATE create $\lambda$ child solutions using perturbation on the parent
    \STATE select the best solution
    \STATE set the best solution as the parent
  \ENDFOR
  \UNTIL{\textit{TerminationCriterion}}
\end{algorithmic}
\end{algorithm}

Using algorithm parameters $\mu$ and $\lambda$, we can balance between the diversification and intensification segments of the search. 
In our experiments, we have used the value of 4 for the parameter $\lambda$, which is a common choice in diverse ES applications, see, e.g.,~\cite{10.1080/03052150500035658,10.1007/978-3-030-16692-2_15,Miller2011}.
With this parameter value, the parent for the next iteration is selected among five solutions in total (the parent and four child solutions).
If multiple parents are used, we set the parameter $\mu$ to the value of 5. Note that we experimented with several values for $\mu$, and the main difference from the performance perspective is in the speed of convergence, realizing in slightly worse coverage in our preliminary experiments (cf. \Cref{app:parameter_selection}).

Let us consider more why taking a small $\lambda$ size (but larger than 1) makes sense. First, if we consider an extreme case where $\lambda$ equals 1, we effectively reach a local search algorithm. While such an algorithm could work for this problem, it would face issues with a high probability of getting stuck in local optima. %
The second extreme for $\lambda$ would represent a large population size (e.g., order of magnitude 100). Then, we face two issues:
\begin{compactitem}
\item Due to the large population size, we must conduct more evaluations~\footnote{Alternatively, we would need to reach good solutions in only a few generations, which is highly unlikely for a problem of such difficulty and the lack of structure in the genotype.}, which will be a problem as fitness evaluation is computationally expensive.
\item When having a large population size, it is also common to use the crossover operator to foster search space exploitation, which increases the computational complexity of the algorithm but also makes tuning more difficult. Indeed, by adding crossover, we must tune the algorithm for different crossover operators and the probability of the crossover action.
\end{compactitem}

Finally, unlike in a classical optimization scenario, here, the algorithm's efficiency is not measured based just on the final, best solution the algorithm has found.
Since the optimization is performed concurrently with the fuzzing, every candidate solution that appears during the algorithm run contributes to the overall fuzzing efficiency.
For that reason, and because the optimization is performed per-target basis, it is important to provide a fast convergence, which can be acquired with a smaller population size.

\subsection{Solution Encoding and Perturbation}%
\label{subsec:encoding}

The ES algorithm can be used with any form of solution encoding, as long as a suitable perturbation operator (or operators) is defined.
In the case of optimizing the fuzzing mutation schedule, we used two solution encodings and corresponding perturbation operators.

First, we investigated an encoding that uses a \textit{real-valued vector} to represent relative probabilities of mutation operators; this representation is equal to the one used in MO{\small PT} \normalsize~\cite{mopt}.
The size of the vector is equal to the number of mutation operators since each element of the vector represents the relative probability that a certain mutation operator (given in Table~\ref{tab:mut_operators}) will be selected.
In each invocation, the values in the vector are used to determine the next mutation operator.
Initial values of vector elements are generated uniformly at random in the range $[0, 1]$.
As the perturbation operator, we use a simple Gaussian perturbation with zero mean and standard deviation of $0.25$; the obtained random value is added to a single randomly selected element in the vector (Figure~\ref{fig:float_encoding}, Appendix~\ref{app:ea}). The value of $0.25$ is selected after tuning, where we followed common reasoning for ES: the operator needs to be able to do significant changes (thus, we do not select a very small standard deviation), but it also should not behave like a random search (which would happen with a large standard deviation value).
The values are always kept greater than zero but are allowed to exceed 1 (to allow the algorithm to emphasize an operator if needed).

The second encoding uses a \textit{binary vector} (with values assuming only 0 and 1), where each element in the vector corresponds to a mutation operator.
This simplifies the mutation operator choice so that only a subset of operators, whose corresponding values in the vector are 1, are used for mutation selection; among the elements of this subset, a random mutation is selected by the fuzzer.
As the perturbation operator, a simple one-bit flip is used; each time a solution needs to be modified, a randomly selected bit in the vector is inverted (Figure~\ref{fig:binary_encoding}, Appendix~\ref{app:ea}).

We decided on the binary encoding for the solution encoding since a preliminary evaluation showed a geometric mean coverage increase of around 3\%. What is more, with the binary encoding, we do not need to tune the standard deviation value for the perturbation operator (as we needed for the real-valued representation). This design decision addresses Challenge~\ref{itm:algo}.

\subsection{Objective Function}

The algorithms described above can be used with any conceivable performance measure related to the process being optimized.
In this case, the primary criterion used for the evaluation of individual solutions is the number of unique paths encountered in the instrumented application.
Unique paths encode all different ways to reach every possible basic block.
While keeping track of all of them is tough (and leads to state explosion), counting new unique paths per iteration is simple and efficient.
Hence, we decided to leverage the number of new unique paths as a feedback signal, especially since most fuzzers already provide this number.
The solution with the highest number of paths will get selected as the next parent.
Thus, our goal is the maximization of the following expression, which is in the evolutionary computation field commonly denoted as the fitness function:

\begin{equation}
    fitness = \# \ Unique\_Paths
\end{equation}

This performance measure follows previous work~\cite{mopt}, but the proposed optimization method can be used to optimize a different criterion if necessary.
An alternative approach to a single criterion would be to use a multi-objective optimization algorithm, but this choice is justified only when conflicting objectives need to be optimized concurrently, which is not the case here. Furthermore, using simpler fitness functions has the advantage of better interpretability, i.e., it is clear why a certain solution is better than some other one.

By combining our simple algorithm design (small population, no need for the user to tweak the parameters), support for various solution's encodings, and fitness function, we address Challenge~\ref{itm:adoption}. 
We emphasize that Evolution Strategy is commonly used in the $(\mu + \lambda)$ form, where standard values are 1 (note that here we talk about the number of parents in a single search, and not the total number of parents due to the parallel execution of ES) and 4, see, e.g.,~\cite{10.1023/A:1015059928466,Hansen2015}. Thus, while one could experiment with other values and then consider $\mu$ and $\lambda$ as parameters that need to be tuned, our investigation shows this is unnecessary. Consequently, we do not consider $\mu$ and $\lambda$ as user parameters, nor would the change of those values result in significant performance differences.

\section{Implementation}
\label{sec:impl}

We implemented a prototype of \coolname in C as an extension to AFL 2.54b~\cite{afl}, a popular generic fuzzer that is leveraged by many research works as a foundation~\cite{mopt,aflgo,redqueen,kAFL}. \changed{\coolname consists of about 320 lines of code.}
AFL is easily extendable and does not contain other algorithmic improvements itself, unlike projects like AFL++~\cite{afl++} that try to incorporate all state-of-the-art improvements for best results in practice. For our \coolname mutation scheduling algorithm, we added an interface to AFL to report feedback in the form of newly discovered paths from the instrumented application to the mutation scheduler. \changed{The interface exposes three functions: initialization, selecting a mutation, and reporting feedback to \coolname.}
This enables a modular design for different mutation scheduling algorithms.

To derive the random numbers needed for our Evolution Strategy, we leverage the RomuDuoJr random number generator (RNG)~\cite{overton2020romu} to balance out the higher reliance on the random number generation of DARWIN's ES algorithm. In \Cref{app:preliminary_experiments}, we show that the speed difference is negligible compared to the standard RNG.

\begin{figure*}[t]
    \centering
    \subfigure[\coolname]{\includegraphics[width=0.32\textwidth]{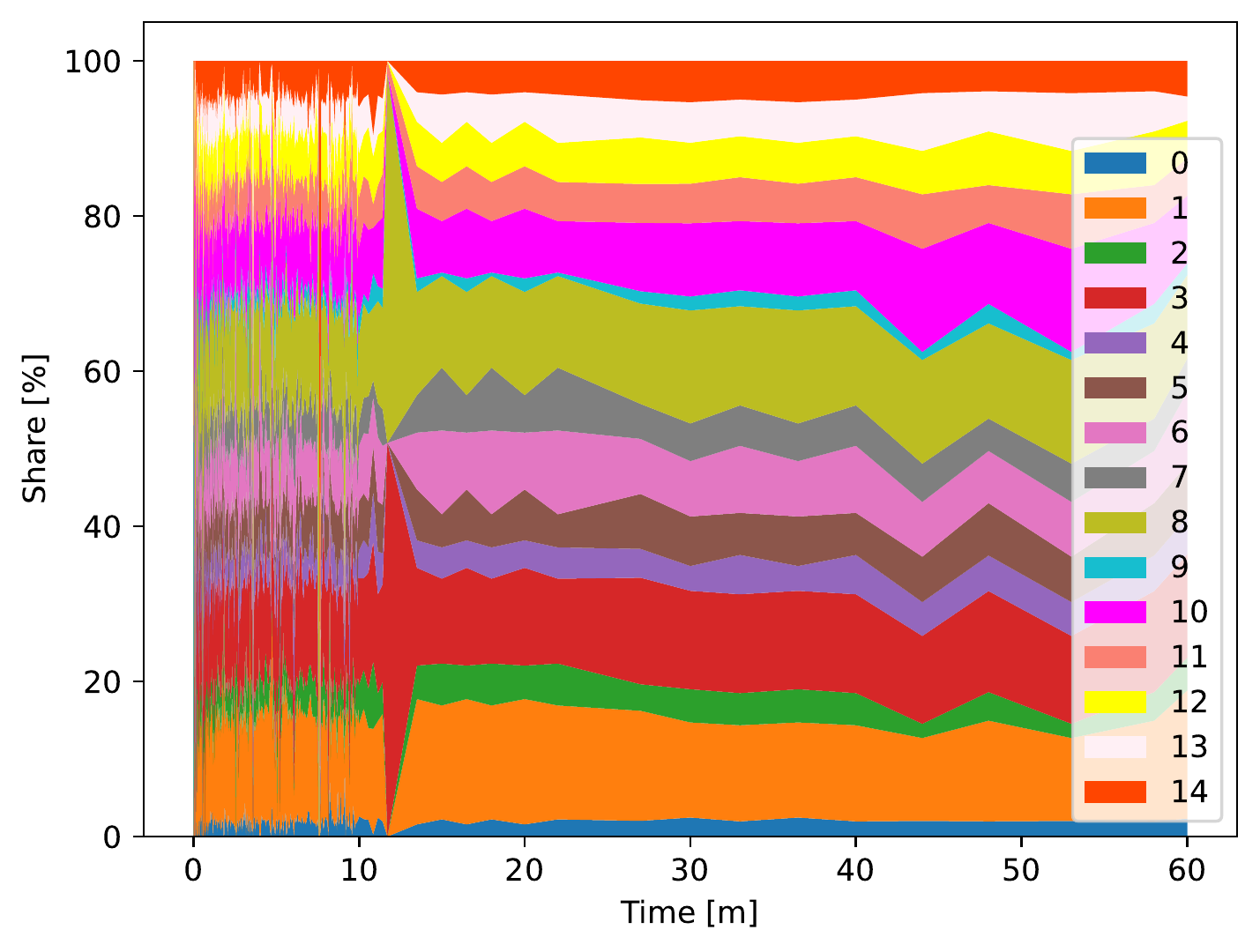}}
    \subfigure[MOPT]{\includegraphics[width=0.32\textwidth]{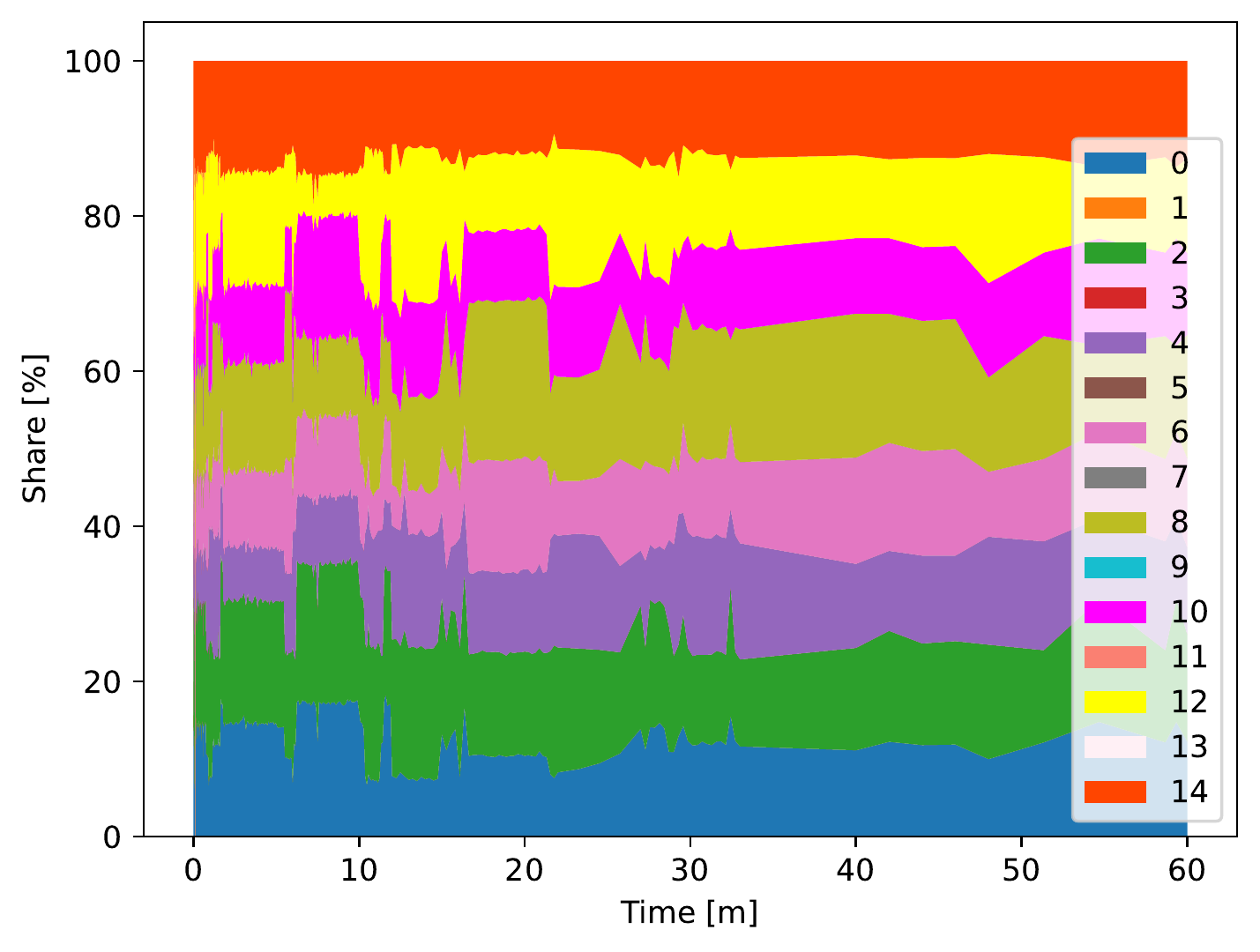}}
    \subfigure[AFL]{\includegraphics[width=0.32\textwidth]{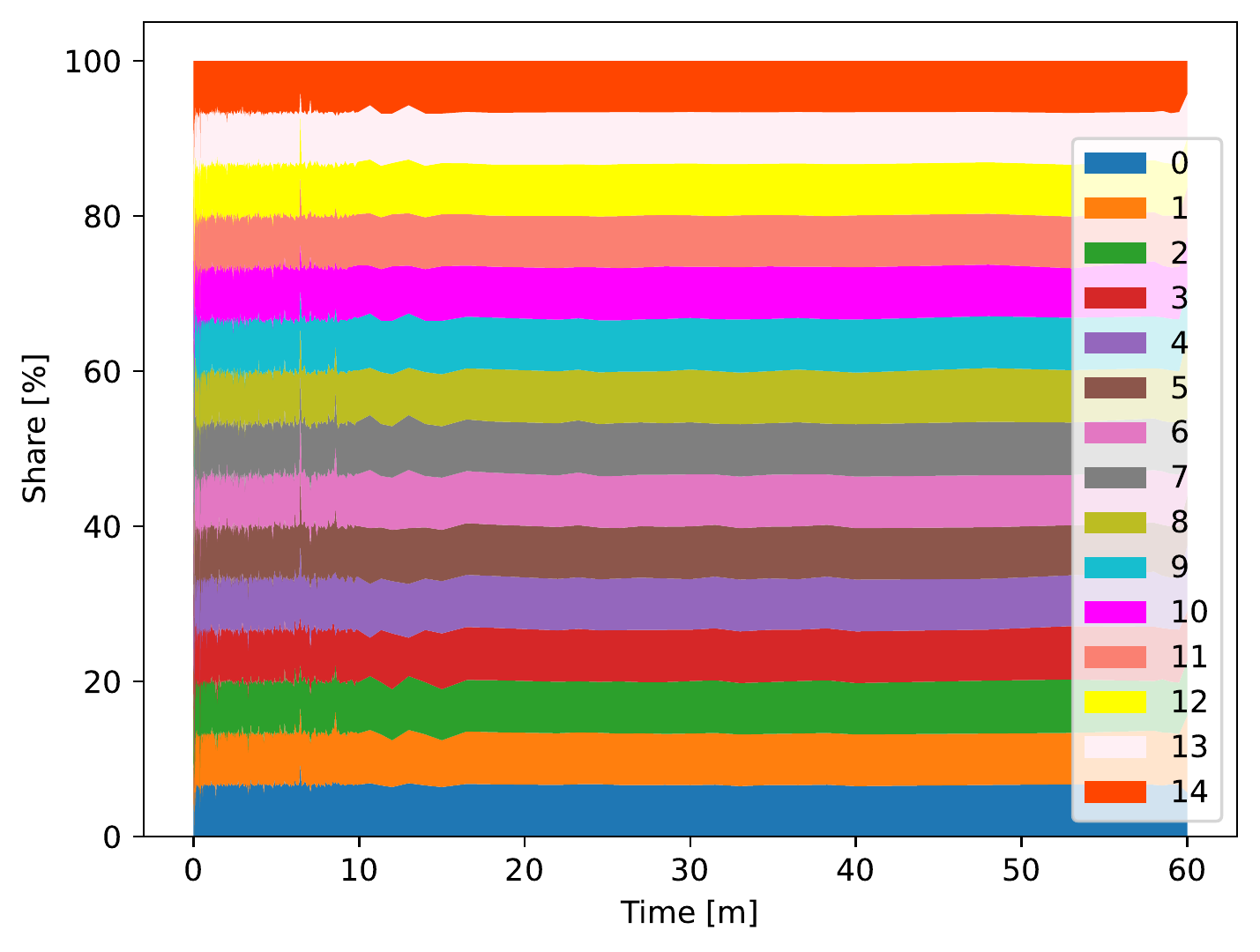}}
    \caption{Mutation history for \texttt{cxxfilt}.}
    \label{fig:mut_history_1}
\end{figure*}

\section{Evaluation}
\label{sec:evaluation}

We analyze \coolname regarding a variety of aspects.
First, we evaluate \coolname's general ability to explore programs as an approximation for the fuzzer's efficiency in~\Cref{subsec:cov}. Second, we evaluate the fuzzers in terms of execution speed in~\Cref{subsec:eval_speed} to ensure our efficiency improvement can be attributed to the novel mutation scheduling algorithm and that the algorithm does not have grave consequences on execution speed.
Finally, we evaluate \coolname's ability to find crashes using the LAVA-M~\cite{dolan2016lava} (\Cref{subsec:eval_lava}) and MAGMA~\cite{magma} (\Cref{subsec:eval_magma}) benchmarks, to show that the aforementioned aspects lead to finding more bugs faster.\\

\textbf{Setup.} Our evaluation setup across all experiments consists of four workstations with an AMD EPYC 7402P 24-Core processor and 256GB of RAM (to perform the evaluation in parallel while keeping memory accesses independent). The target applications, fuzzers, and seeds are all stored on a ramdisk to reduce the influence of disk I/O.
Each evaluation run is executed sequentially on a dedicated machine to reduce the influence of, e.g., memory bandwidth.

We evaluate \coolname against the most-related work, MO{\small PT}\normalsize{}, and AFL 2.54b as a baseline (as both \coolname and MO{\small PT}\normalsize{} extend AFL).
We ported MO{\small PT} \normalsize to AFL 2.54b (by diffing AFL 2.52b and AFL 2.54b) to ensure that MO{\small PT} \normalsize got the same bug fixes that \coolname and AFL have.

Evaluation of fuzzers is, as with most research topics in security, not standardized, leading to fluctuating results reported in papers and varying results in practice. This is mainly due to two aspects: 1) evaluating related work with non-optimal parameters and 2) missing statistical analysis of the results. 
For the former, we disable the deterministic stage of \coolname and AFL for all experiments completely while using the corresponding Pacemaker mode (with the parameter ``-L 0'' ) to achieve the same effect and focus on the havoc stage for MO{\small PT}\normalsize.
Note that this is crucial for a fair comparison~\cite{wu2022one}.
For the latter, we integrated the approaches proposed by Klees et al.~\cite{klees2018evaluating} to the best of our knowledge and investigated broadly used fuzzing benchmarks to reason about \coolname's performance.

\subsection{Evaluating Coverage}
\label{subsec:cov}

\begin{figure*}[t!]
    \centering
    \subfigure[\texttt{bsdtar}]{\includegraphics[width=0.43\textwidth]{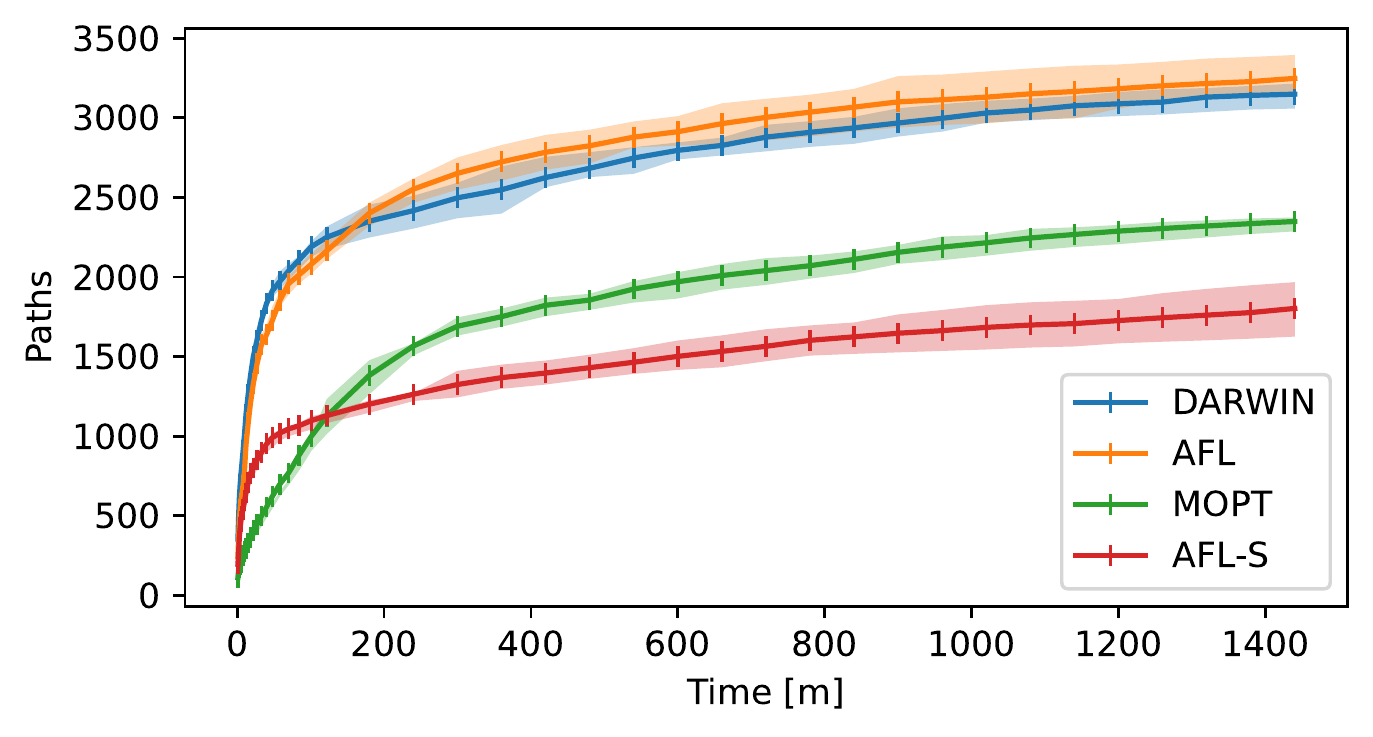}} 
    \subfigure[\texttt{cxxfilt}]{\includegraphics[width=0.43\textwidth]{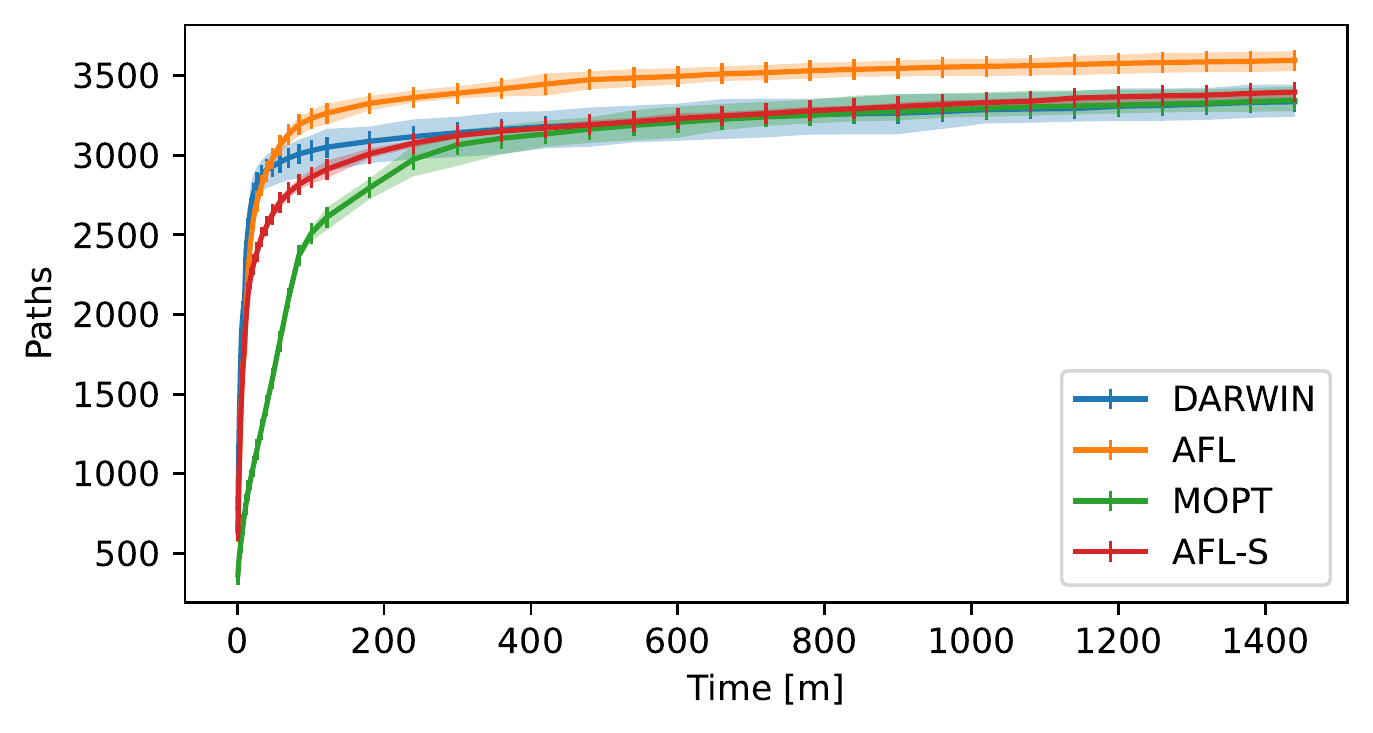}} 
    \subfigure[\texttt{djpeg}]{\includegraphics[width=0.43\textwidth]{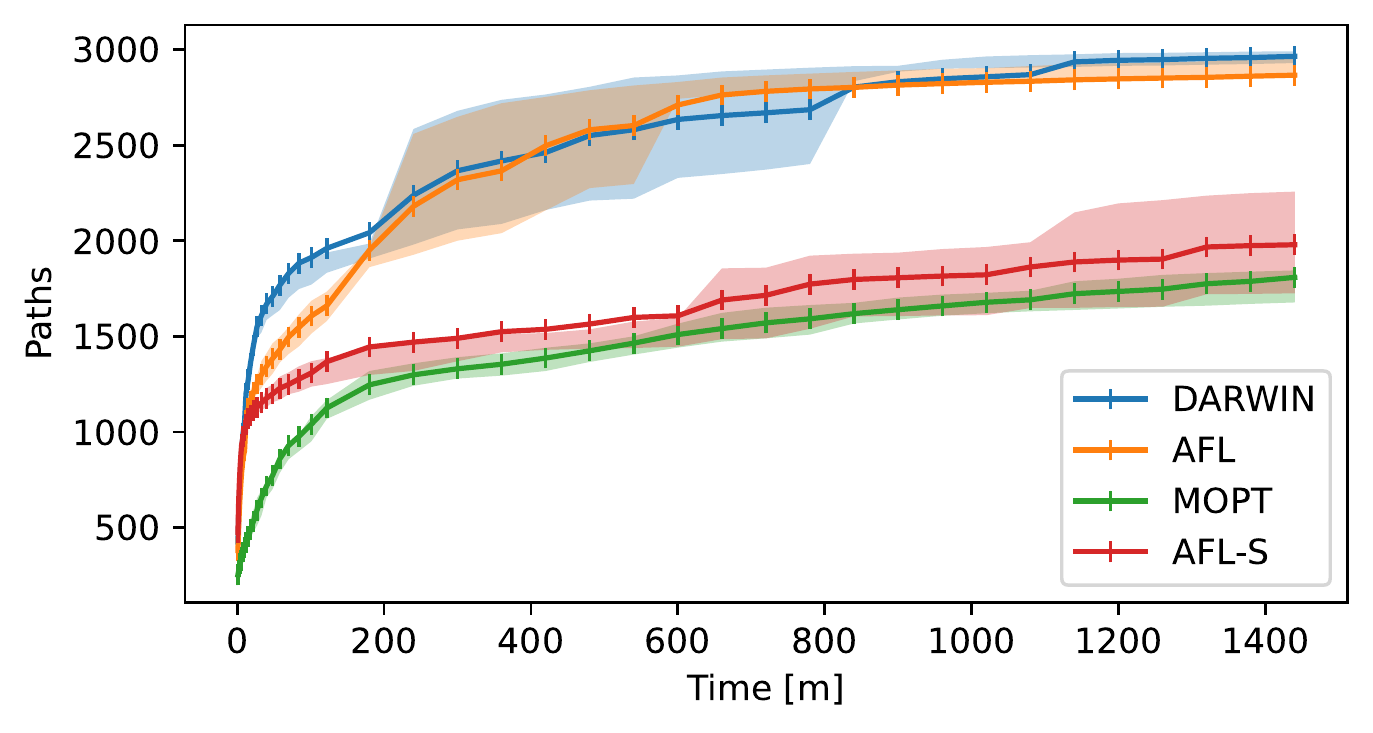}}
    \subfigure[\texttt{jhead}]{\includegraphics[width=0.43\textwidth]{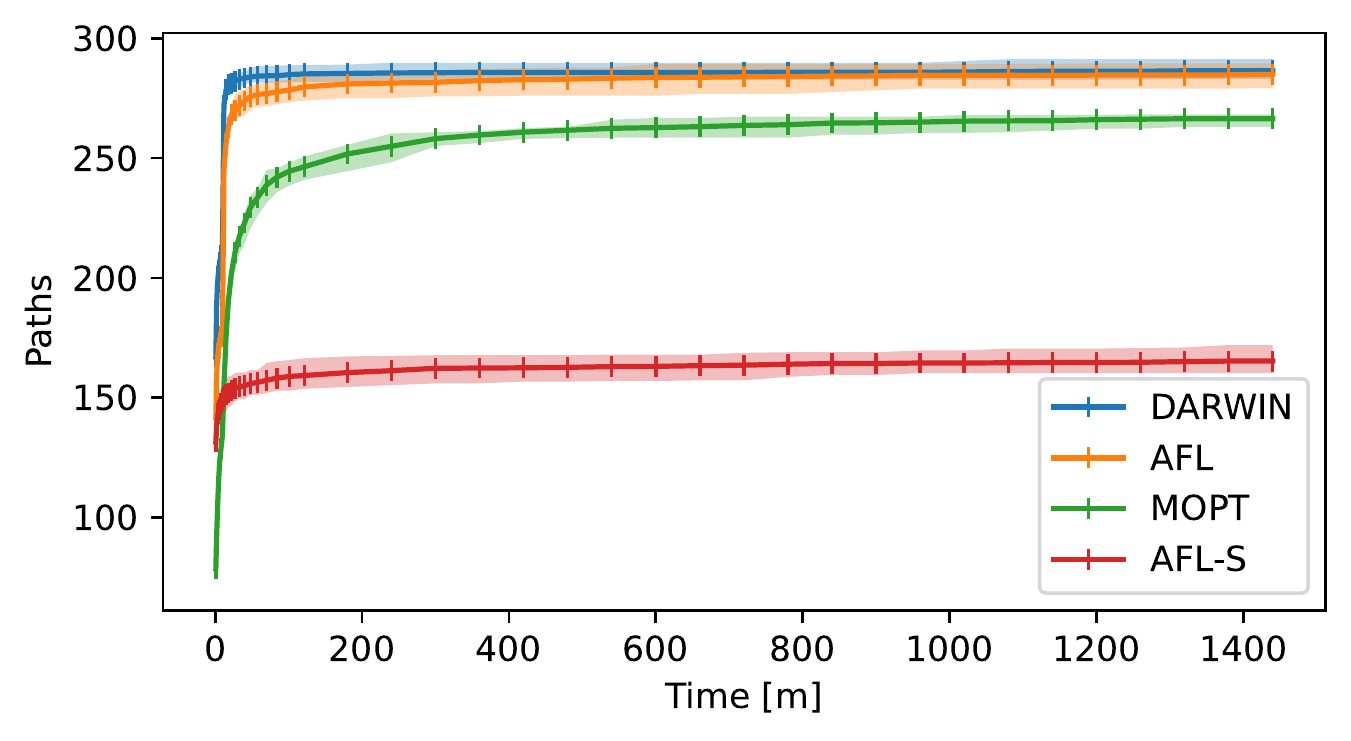}}
    \subfigure[\texttt{objcopy}]{\includegraphics[width=0.43\textwidth]{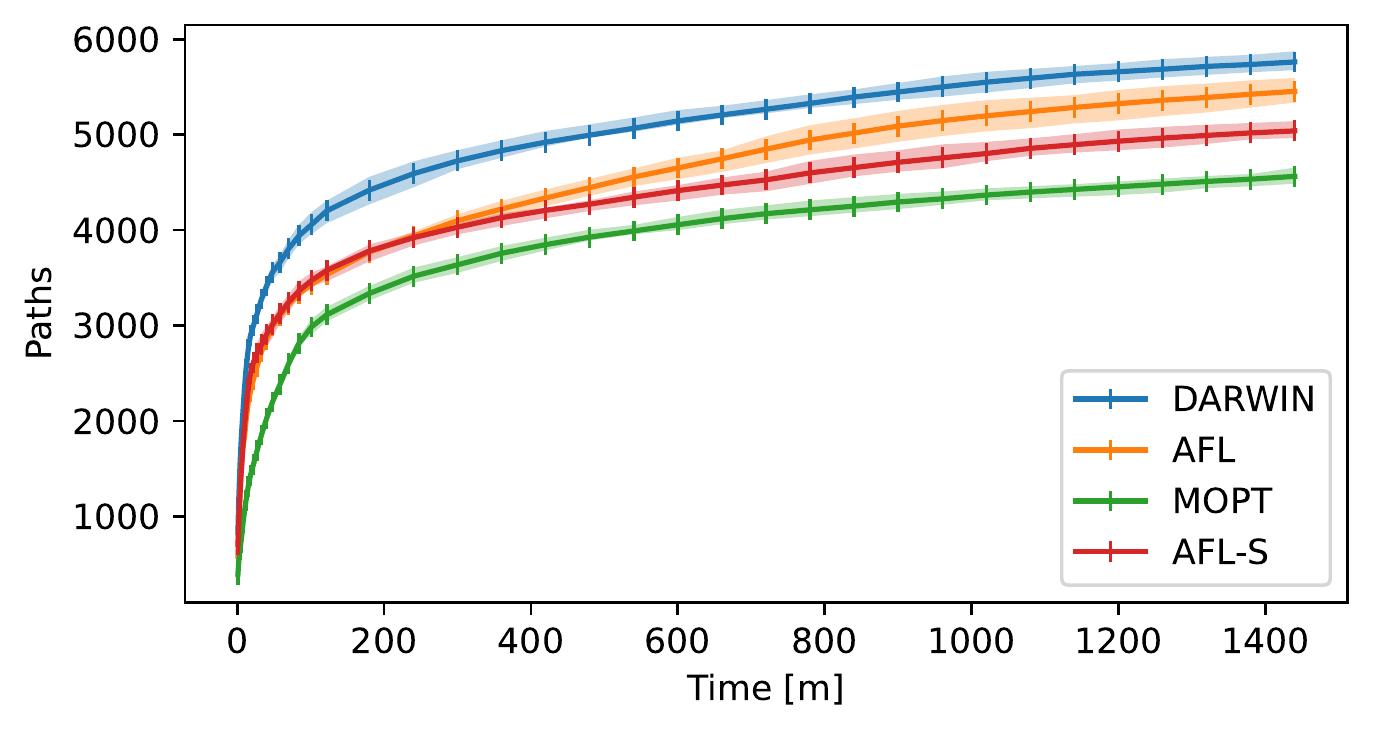}}
    \subfigure[\texttt{objdump}]{\includegraphics[width=0.43\textwidth]{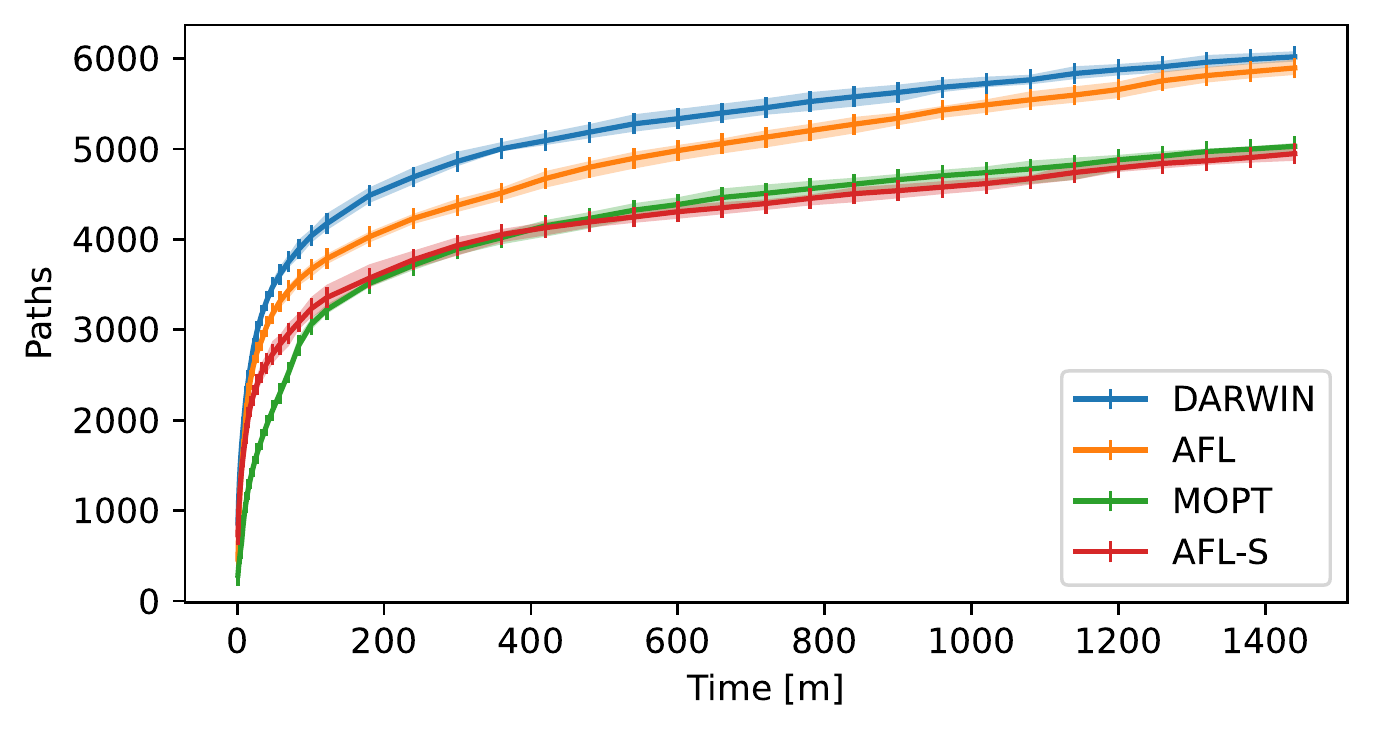}}
    \caption{The coverage results on the various benchmarks for AFL, MOPT, \coolname, and the statically optimized variant AFL-S. Shaded areas represent the respective 25\%/75\% quartiles.}
    \label{fig:cov1}
\end{figure*}

In the first step, we use code coverage as a proxy metric for a fuzzer's success. While code coverage is a well-established quality measure in related work~\cite{mopt,ecofuzz,aflgo,redqueen}, it merely approximates the fuzzer's capabilities in finding bugs, as a fuzzer needs to cover a line of code to find a bug in it.

In all experiments, we leverage six applications, which process an executable ELF file without modifying it, from the well-fuzzed GNU binutils suite~\footnote{\url{https://www.gnu.org/software/binutils/}} in version 2.34~\cite{redqueen,ecofuzz,mopt,fairfuzz}. We further include jhead 3.06.0.1, bsdtar (from libarchive) 3.6.0, tcpdump 4.99.1, and djpeg 2.1.2, as they are also commonly used~\cite{ecofuzz,redqueen,mopt,fairfuzz}. For increased reproducibility, we also kept the number of seed files low. Otherwise, as each seed is selected randomly by default, the variance for each run increases. The seeds used for the binutils targets always remain the same: one uninformed, empty test case and one minimal correct test case. We used the standard testcases bundled with AFL, except for binutils, where we used a minimal C program (smaller than the one bundled) described in~\Cref{app:binutils_seed}. \\

\begin{table}[]
    \caption{Invocation of benchmark tools and file formats used as seeds.}
    \centering
    \begin{tabular}{|l|l|c|}
    \hline
    Benchmark & Invocation                                 & Format \\\hline
    bsdtar    & -xf @@ /dev/null                           & TAR    \\
    cxxfilt   & -t                                         & ELF    \\
    djpeg     & @@                                         & JPEG   \\
    jhead     & @@                                         & JPEG   \\
    objcopy   & --dump-section text=/dev/null @@ /dev/null & ELF    \\
    objdump   & -d @@                                      & ELF    \\
    readelf   & -a @@                                      & ELF    \\
    size      & @@                                         & ELF    \\
    strip     & -o /dev/null @@                            & ELF    \\
    tcpdump   & -nr @@                                     & PCAP  \\ \hline
    \end{tabular}
    \label{table:invocation}
\end{table}

We evaluate the performance of the selected fuzzers over three independent runs, reporting the mean and 25\%/75\% quartiles. Each experiment runs for 24 hours.
We present the mean coverage for each benchmark but also the standard deviation over time. Further, we additionally conduct the non-parametric Mann-Whitney U test to evaluate whether there are statistically significant differences among results, as suggested by Arcuri et al.~\cite{arcuri2014hitchhiker} and Klees et al.~\cite{klees2018evaluating}.\\\\

\begin{figure*}[t!]
    \centering
    \subfigure[\texttt{readelf}]{\includegraphics[width=0.43\textwidth]{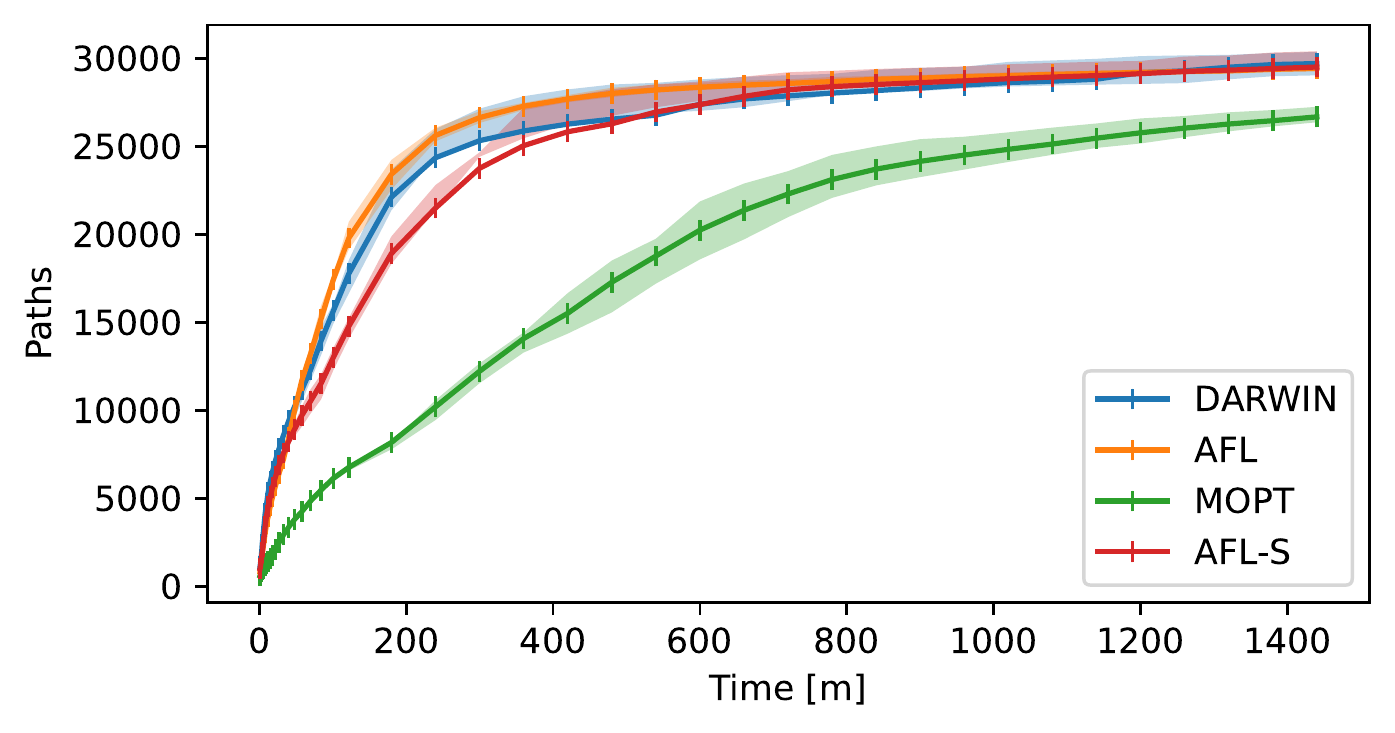}}
    \subfigure[\texttt{size}]{\includegraphics[width=0.43\textwidth]{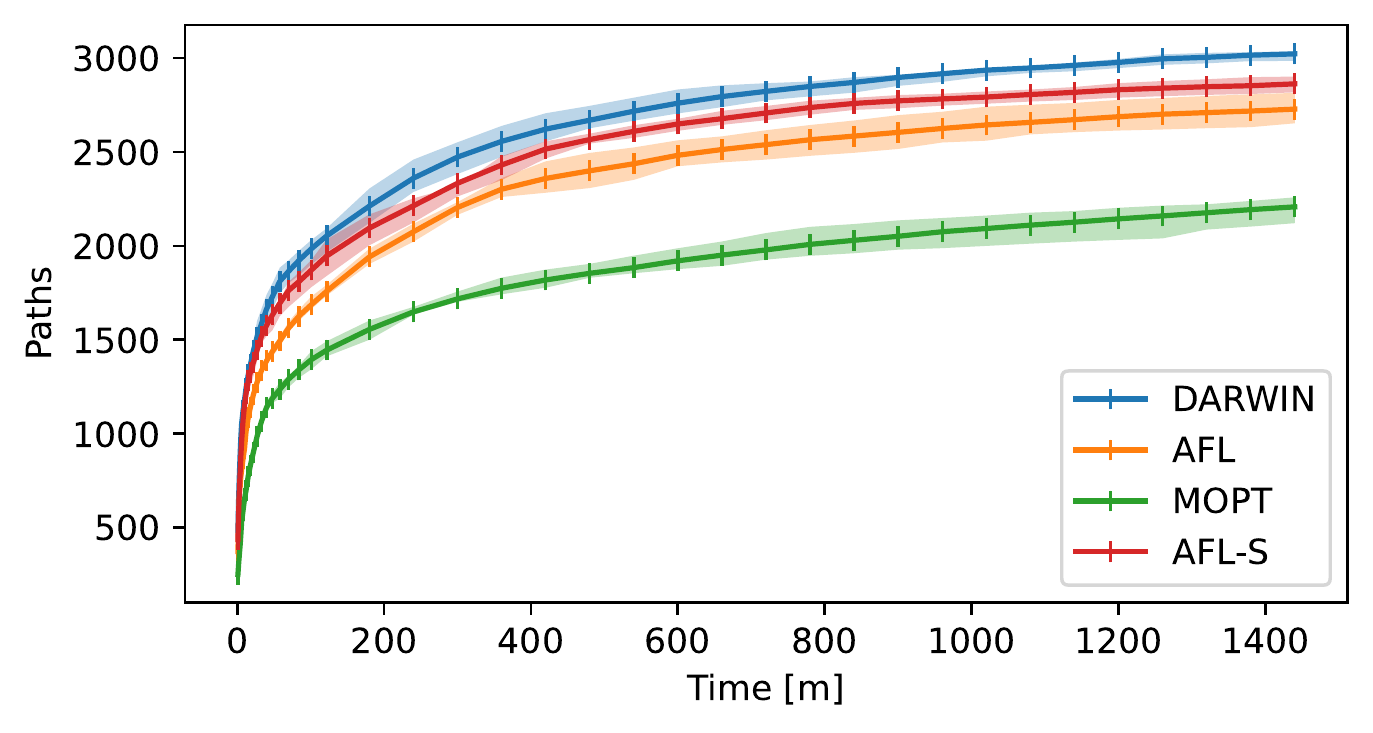}} 
    \subfigure[\texttt{strip}]{\includegraphics[width=0.43\textwidth]{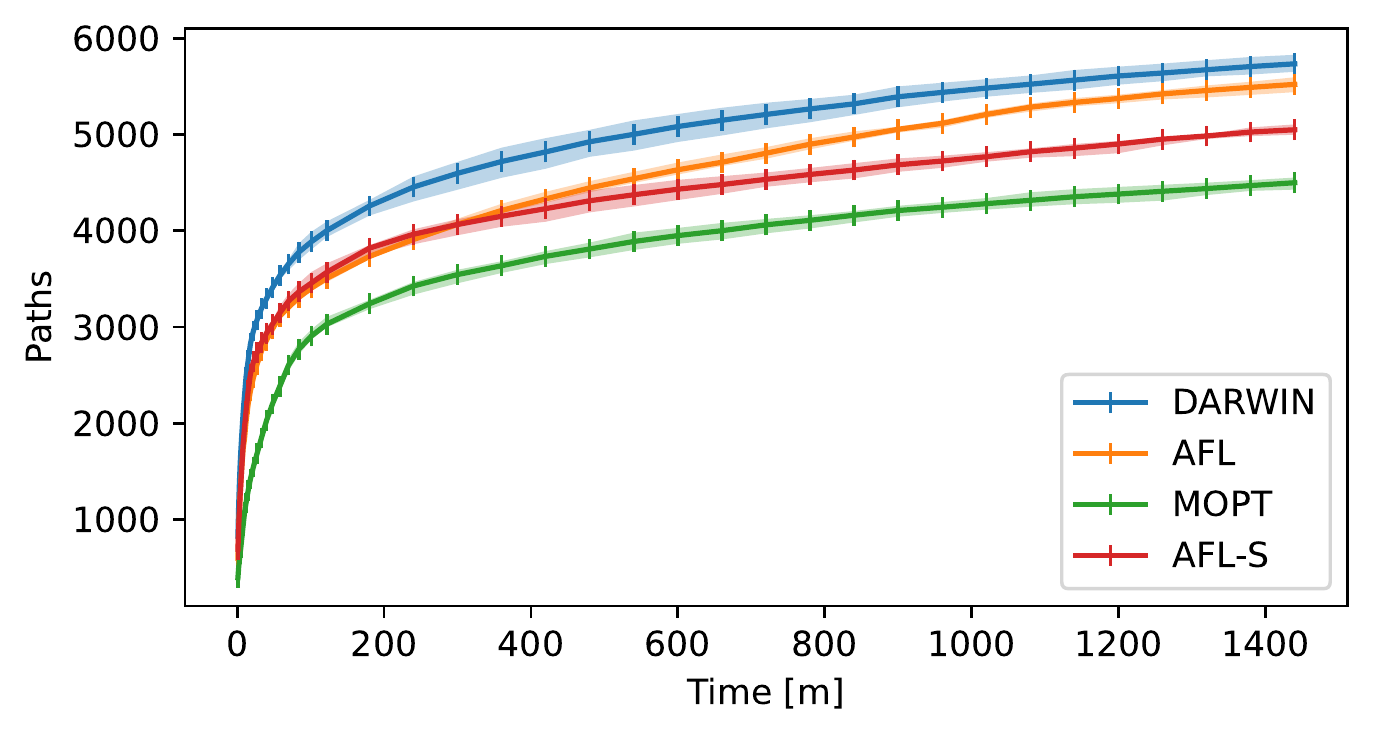}}
    \subfigure[\texttt{tcpdump}]{\includegraphics[width=0.43\textwidth]{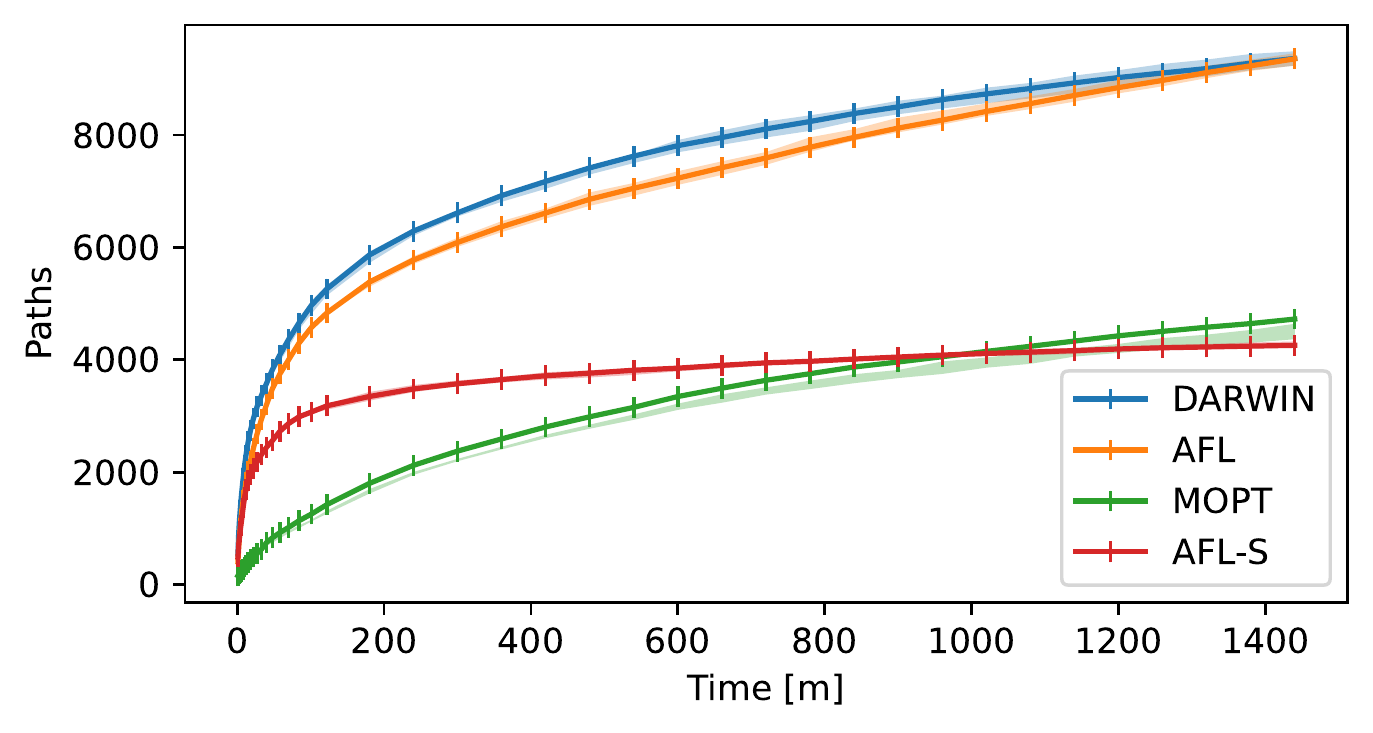}}
    \caption{The coverage results on various benchmarks for AFL, MOPT, \coolname, and the statically optimized variant AFL-S. Shaded areas represent the respective 25\%/75\% quartiles.}
    \label{fig:cov1_2}
\end{figure*}

The results of our coverage evaluation for \coolname, MO{\small PT}\normalsize{}, and AFL are depicted in~\Cref{tab:eval_coverage}.
In~\Cref{fig:cov1} and~\Cref{fig:cov1_2}, we show the respective graphs for coverage over time.

\begin{table*}[ht!]
    \centering
    \caption{Mean coverage results measured in unique paths and edges for well-fuzzed targets over ten runs. AFL-S is AFL with optimized, static probability distribution. Geometric mean improvement (``geomean'') of \coolname over MOPT and AFL, respectively. p-values for the Mann-Whitney U test for \coolname on the number of unique paths found in 24h. p-values for the Mann-Whitney U test for \coolname on the number of unique paths found in 24h. Note that experiments with a similar result across samples (italic) lead to a high p-value naturally; all remaining experiments are  statistically significant with $ p < 0.05$.}
    \scriptsize
    \begin{tabular}{|l|rr|rrr|rrr|rrr|}
        \hline
        & \multicolumn{2}{c|}{\coolname}& \multicolumn{3}{c|}{MOPT}& \multicolumn{3}{c|}{AFL} & \multicolumn{3}{c|}{AFL-S}\\
        Benchmark & unique paths & edges & unique paths & edges & p-value & unique paths & edges & p-value & unique paths & edges & p-value \\ \hline
        \texttt{bsdtar} & 3147.20 & \textbf{5369.70} & 2347.50 & 4832.0 & 9.13e-05 & \textbf{3246.50} & 5302.60 & \textit{0.093} & 1801.30 & 4970.90 & 9.08e-05 \\
        \texttt{cxxfilt} & 3334.18 & 2327.27 & 3343.00 & 2333.09 & 0.0001 & \textbf{3594.91} & 2425.36 & 1.95e-04 & 3395.50 & \textbf{2500.30} & \textit{0.647} \\
        \texttt{djpeg} & \textbf{2964.60} & \textbf{3191.00} & 1807.80 & 2765.90 & 9.13e-05 & 2866.00 & 3148.80 & \textit{0.163} & 1978.50 & 2851.80 & 9.13e-05 \\
        \texttt{jhead} & \textbf{285.40} & \textbf{340.00} & 265.4 & 339.00 & 2.17e-04 & 283.90 & \textbf{340.00} & \textit{0.520} & 164.30 & 336.00 & 8.88e-05 \\
        \texttt{objcopy} & \textbf{5760.82} & \textbf{7912.36} & 4562.00 & 7606.00 & 4.08e-05 & 5453.09 & 7881.27 & 4.05e-04 & 5038.20 & 7507.90 & 6.20e-05 \\
        \texttt{objdump} & \textbf{6018.91} & \textbf{7269.82 }& 5028.82 & 7003.73 & 4.06e-05 & 5895.91 & 7141.55 & 0.028 & 4947.90 & 7044.00 & 6.20e-05 \\
        \texttt{readelf} & \textbf{29715.64} & 13012.36 & 26686.73 & 12273.00 & 4.08e-05 & 29439.27 & 12032.18 & \textit{0.162} & 29519.90 & \textbf{13019.20} & \textit{0.805} \\
        \texttt{size} & \textbf{3020.91} & \textbf{4030.91} & 2206.82 & 3773.45 & 4.07e-05 & 2726.91 & 3809.55 & 5.32e-05 & 2861.50 & 3941.00 & 8.24e-04 \\
        \texttt{strip} & \textbf{5732.55} & 7703.55 & 4497.36 & 7470.82 & 4.08e-05 & 5519.36 & \textbf{7756.45} & 0.001 & 5047.60 & 7354.30 & 6.20e-05 \\
        \texttt{tcpdump} & \textbf{9361.20} & \textbf{13834.10} & 4723.60 & 11618.70 & 9.13e-05 & 9354.10 & 13317.2 & \textit{1.0} & 4255.70 & 11952.10 & 9.13e-05 \\ \hline
        geomean &       &       & +29.40\% & +6.77\% &       & +1.60\% & +1.73\% &       & +32.35\% & +4.38\% &  \\
        \hline
    \end{tabular}
    \label{tab:eval_coverage}
\end{table*}

\begin{figure*}[t]
    \centering
    \subfigure[\coolname]{\includegraphics[width=0.32\textwidth]{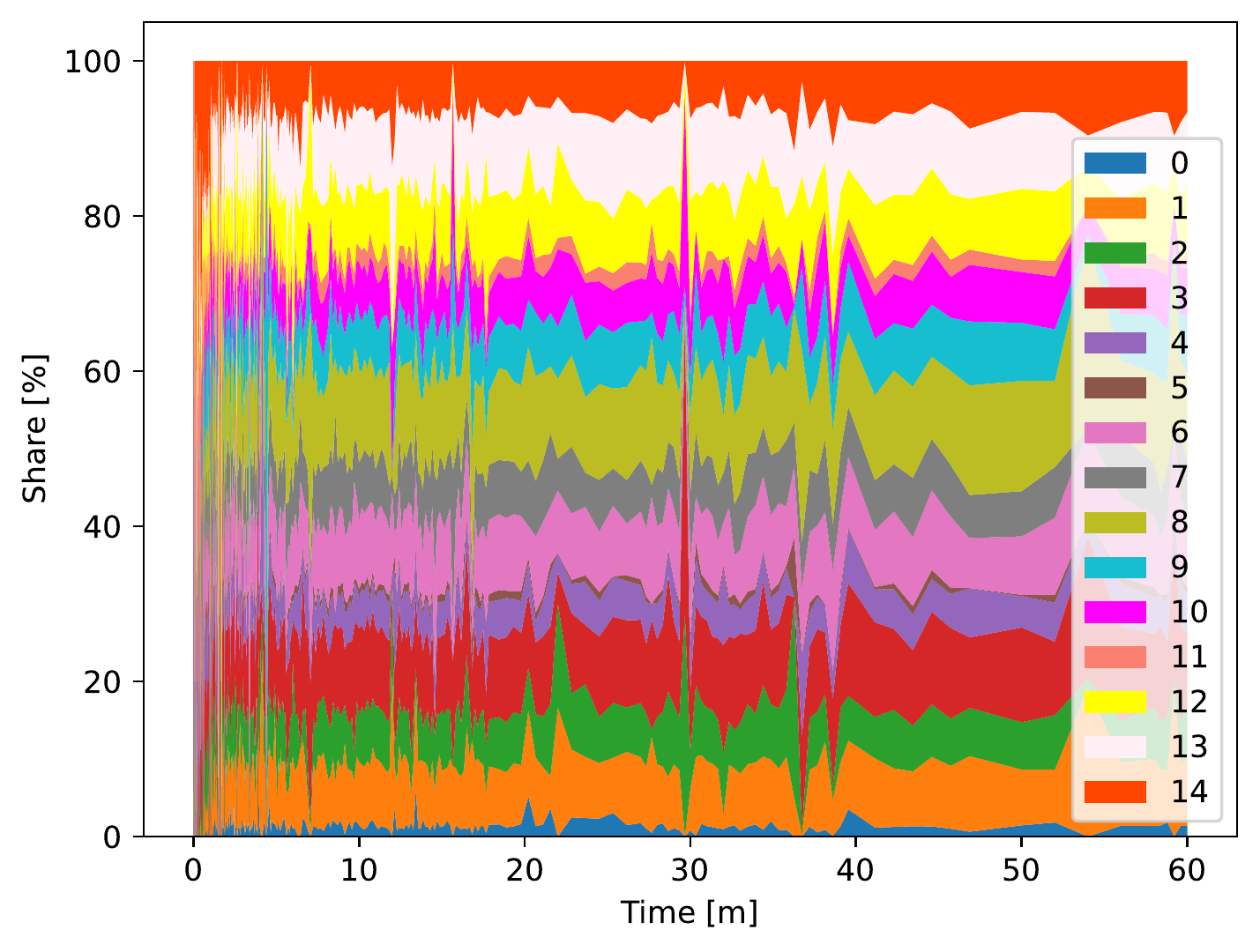}}
    \subfigure[MOPT]{\includegraphics[width=0.32\textwidth]{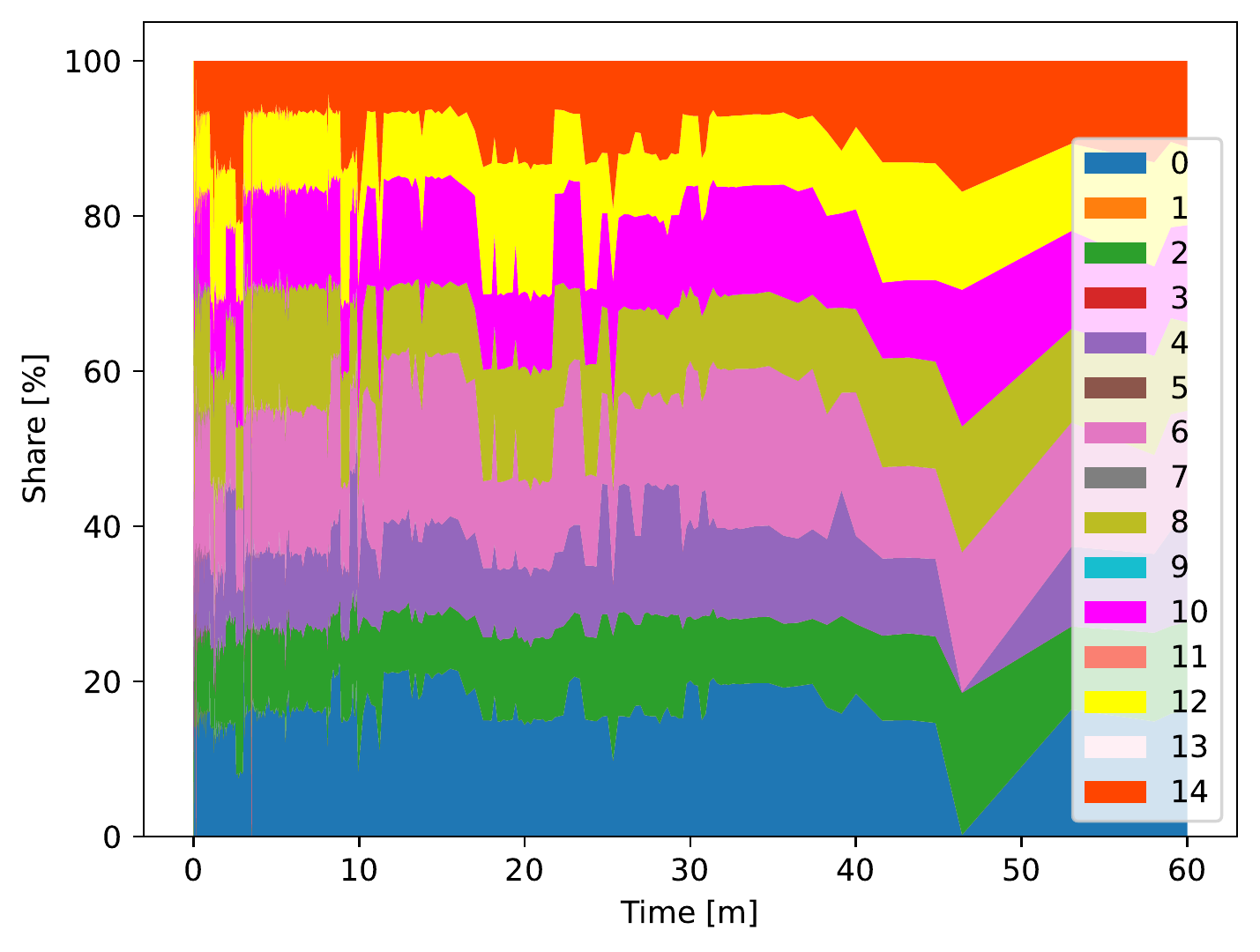}}
    \subfigure[AFL]{\includegraphics[width=0.32\textwidth]{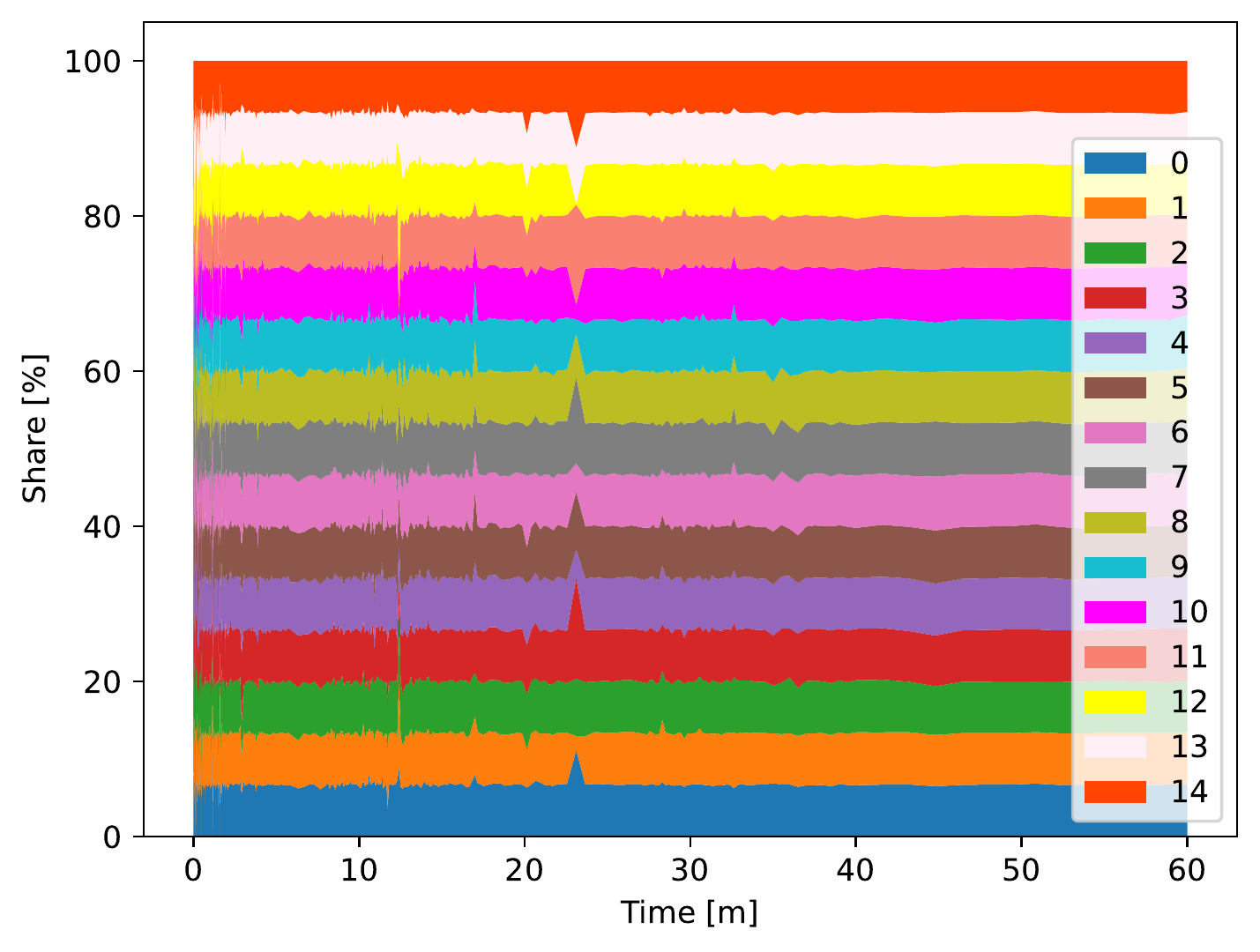}}
    \caption{Mutation history for \texttt{size}.}
    \label{fig:mut_history_2}
\end{figure*}

First of all, we can observe that MO\small{PT} \normalsize is constantly performing worse than \coolname, as well as AFL (except in one experiment).
For \texttt{djpeg}, \texttt{jhead}, \texttt{objcopy}, \texttt{objdump}, \texttt{size}, and \texttt{tcpdump}, \coolname clearly reaches the highest number of paths and edges, and also has the steepest increase in unique paths found over time for the first hour of fuzzing.
For \texttt{objcopy} and \texttt{strip}, we saw that \coolname the probability for mutation 0 (flip single bit) and 14 (overwrite bytes with a randomly selected chunk) tremendously, whereas \coolname reduces probability of mutation 4 (randomly subtract from byte) for \texttt{objdump} and 5 (randomly add to byte) for \texttt{size}.
Besides looking only at the paths covered, we can also consider the time to the same coverage as a figure of merit.
For example, for \texttt{size}, \coolname reaches AFL's maximum coverage approximately 800 minutes earlier, similar for \texttt{objcopy} and \texttt{objdump} where \coolname reaches the same point approx. 700 minutes earlier. 

The \texttt{cxxfilt} benchmark shows such a different behavior than other benchmarks that it warrants further discussion. This is the only case where AFL is a clear winner, and both mutation-scheduling-based fuzzers reach a similar coverage.
While we noticed that AFL is achieving new coverage with the splicing stage around 50\% more often than \coolname, MO{\small PT}\normalsize{} found four times as many coverage-triggering inputs using splicing.
As such, we can exclude splicing being one reason for this effect.

Hence, we looked at the mutations scheduled within a timespan of 1h, as shown in~\Cref{fig:mut_history_1}.
There we can see that \coolname as well as MO{\small PT}\normalsize{} put more and more emphasis on mutators 8 and 10 after around 40 minutes.
This is also the very same moment where AFL starts to outperform both fuzzers.
As \texttt{cxxfilt} is aiming at demangling overloaded functions (and the similarly behaving bsdtar is unpacking archives), it seems like mutation schedulers only add little benefit to fuzzing targets that are heavily relying on parsing. Yet, their performance impact (as we explore later) reduces the raw execution speed of the fuzzer, resulting in inferior coverage results.

Looking at \texttt{size}~\Cref{fig:mut_history_2}, a target where \coolname significantly outperforms AFL and MO{\small PT}\normalsize{}, \coolname avoids scheduling mutators 0, 5, 11, while mutator 0 has a large share in MO{\small PT}\normalsize{}.

While analyzing the mutation histories, we noticed that MO{\small PT}\normalsize{}  schedules only 8 of the 15 mutations across all of our benchmarks.
Most likely, this is an implementation bug as there is no visible calibration effect (in comparison to, e.g., the first 10 minutes of \coolname, where \coolname converges quickly afterward).
This is also one possible factor for the diverse results MO{\small PT}\normalsize{} shows in our experiments.

In conclusion, \coolname shows a geometric mean improvement in edge coverage of 6.77\% over MO{\small PT}\normalsize{}, and 1.73\% over AFL, hence, this addresses Challenge~\ref{itm:optimal}.
While this might seem insignificant at first, coverage measurements are only an approximation of a fuzzer's efficiency in finding bugs, as we show later.

\begin{table}[!ht]
    \centering
    \caption{Median relative code-coverages on each benchmark after 10 runs with 6h each. Median relative performance of each fuzzer to the encountered experiment maximum.}
    \begin{tabular}{|l|c|c|c|}
    \hline
         & \textbf{\coolname} & \textbf{AFL} & \textbf{MOPT} \\ \hline
        \textbf{FuzzerMedian} & \textbf{97.11} & 96.89 & 86.70 \\ \hline
        \textbf{FuzzerMean} & \textbf{96.34} & 95.46 & 83.64 \\ \Xhline{2\arrayrulewidth}
        bloaty\_fuzz\_target & \textbf{96.40} & 94.95 & 89.62 \\ \hline
        curl\_curl\_fuzzer\_http & \textbf{98.35} & 97.25 & 92.19 \\ \hline
        freetype2-2017 & \textbf{94.80} & 93.68 & 78.74 \\ \hline
        harfbuzz-1.3.2 &\textbf{ 98.95} & 97.70 & 86.48 \\ \hline
        libjpeg-turbo-07-2017 & \textbf{88.81} & 88.72 & 69.20 \\ \hline
        libpng-1.2.56 & \textbf{99.72} & 98.79 & 94.05 \\ \hline
        libxml2-v2.9.2 & 93.34 & \textbf{96.89} & 61.79 \\ \hline
        libxslt\_xpath & \textbf{97.11} & 92.21 & 83.69 \\ \hline
        mbedtls\_fuzz\_dtlsclient & \textbf{98.94} & 97.63 & 95.26 \\ \hline
        openssl\_x509 & \textbf{99.87} & \textbf{99.88} & 99.73 \\ \hline
        openthread-2019-12-23 & 88.74 & \textbf{88.84} & 86.70 \\ \hline
        php\_php-fuzz-parser & 96.78 & \textbf{98.94} & 94.64 \\ \hline
        proj4-2017-08-14 &\textbf{ 94.95} & 93.40 & 28.57 \\ \hline
        re2-2014-12-09 & \textbf{98.45} & 98.34 & 83.51 \\ \hline
        sqlite3\_ossfuzz & \textbf{92.44} & 86.38 & 78.14 \\ \hline
        systemd\_fuzz-link-parser & \textbf{99.92} & 99.84 & 97.97 \\ \hline
        vorbis-2017-12-11 & \textbf{97.01} & 96.77 & 84.87 \\ \hline
        woff2-2016-05-06 &\textbf{ 97.78} & 95.75 & 91.88 \\ \hline
        zlib\_zlib\_uncompress\_fuzzer & \textbf{98.12} & 97.71 & 92.14 \\ \hline
    \end{tabular}
    
    \label{tab:eval_fuzzbench}
\end{table}

\paragraph{FuzzBench}
FuzzBench~\cite{fuzzbench} is a fuzzing benchmark suite developed by Google. The benchmark comprises various widely-fuzzed real-world targets, e.g., from OSS-Fuzz~\cite{google-ossfuzz}. 
We conducted a local FuzzBench coverage experiment over ten runs, where each run took six hours. 
All Fuzzbench experiments were conducted on a workstation with an Intel Xeon Silver 4110 CPU with 2.10GHz and 128GB RAM.

Experiments are depicted in~\Cref{tab:eval_fuzzbench}.
\coolname outperforms both AFL and MO{\small PT}\normalsize{} in the avg. normalized score and avg. rank. Specifically, \coolname reaches the highest median relative code coverage in 15 out of 19 experiments, is even with AFL in two (\coolname has in \texttt{openssl\_x509} 0.00001\% and in \texttt{openthread-2019-12-23} -0.11\% less coverage).

In the remaining two experiments, AFL slightly outperforms \coolname: \texttt{libxml2-v2.9.2} (3.80\%) and \texttt{php\_php-fuzz-parser} (2.23\%) are both parsers, as such, coverage mainly comes from well-structured testcases.
As \coolname does not improve testcase generation itself, e.g., using grammars, both fuzzers generate testcases of similar (bad) quality and hence, largely fail to cover a big part of the targets. AFL's faster execution speed allows it to generate more testcases per second, which is the cause for the differences.

MO{\small PT}\normalsize{} is last in every experiment, with \texttt{openssl\_x509} being the experiment closest to DARWIN and AFL. 
Thus, \coolname is the first mutation scheduler to show coverage improvements over AFL in FuzzBench.

\paragraph{Static Optimization vs. Adaptive Optimization}
For the mutation scheduling problem at hand, it is not clear if the perfect mutation probability distribution changes over time with the same target application. Hence, we used \coolname to fuzz the targets from~\Cref{subsec:cov} for 24h, but this time, storing the "best so far" parent in the current set of parents after 24h. As shown in~\Cref{tab:eval_coverage}, \coolname outperforms the static variant (referred to as AFL-S) by 4.38\% geometric mean in the number of covered edges (and 32.35\% in paths). Especially in the non-binutils experiments, \coolname shows the importance of adaptive optimization throughout the fuzzing process.

In binutils, the static variant is much closer to the adaptive variant,as a lot of library code is shared between the individual applications, and the inputs are always executables. This also reflects in the resulting probability distributions, i.e., \texttt{readelf}, \texttt{size}, and \texttt{cxxfilt} share the same distribution, and \texttt{strip}, \texttt{objcopy}, and \texttt{objdump} share the same distribution. Both groups have 7 disabled mutations and commonly disable mutations 3, 5, and 8 (cf. \Cref{app:preliminary_experiments}). From our investigations, the mutations left are enough to overcome the initial parsing steps and then concentrate on common library code, which is also what we expect the probability distribution to converge to in later phases in the adaptive variant.

\changed{In all experiments, \coolname outperformed AFL-S also after 200 minutes. In the six experiments where AFL-S eventually reached the same (average) coverage \coolname reached after 200 minutes, it took AFL-S 285 more minutes on average. Further, in four experiments, AFL-S never even reached that mark.}

\subsection{Parameter Selection}
\label{app:parameter_selection}
Even though the parameters for $\mu$ and $\lambda$ are widely consistent throughout literature~\cite{10.1023/A:1015059928466,Hansen2015}, we also evaluated neighboring configurations, as shown in~\Cref{tab:eval_parameter}. Our 24h experiments over ten runs show that within the large body of coverage evaluation targets, the initial configuration still outperforms them.

\begin{table*}[!ht]
    \centering
    \caption{Coverage results measured in unique paths and edges for well-fuzzed targets in binutils over 10 runs, 24h each.}
    \begin{tabular}{|l|r|r|r|r|r|r|r|r|r|r|}
    \hline
                 & \multicolumn{2}{c|}{\coolname ($\mu$:5 $\lambda$:4)}& \multicolumn{2}{c}{$\mu$:5 $\lambda$:3}& \multicolumn{2}{|c|}{$\mu$:5 $\lambda$:5} & \multicolumn{2}{c|}{$\mu$:6 $\lambda$:4} & \multicolumn{2}{c|}{$\mu$:4 $\lambda$:4}\\
    \hline
         Benchmark   &   unique paths &   edges &  unique paths &   edges &  unique paths &   edges &  unique paths &   edges &  unique paths &   edges \\
    \hline
        \texttt{cxxfilt}     & 3334.18 & 2327.27         & \textbf{3375.10}  & \textbf{2365.70}        & 3233.50 & 2301.00     & 3251.70 & 2323.00      & 3224.80 & 2294.10\\
        \texttt{objcopy}     & \textbf{5760.82}   & \textbf{7912.36}       & 5567.10 &  7866.10        &  5564.70 & 7835.60    & 5584.20 & 7821.60      & 5565.10     &   7813.90\\
        \texttt{objdump}     & \textbf{6018.91}  & \textbf{7269.82}        &  5832.80 & 7239.20        & 5820.00  &  7256.00   & 5774.30 & 7221.60      & 5880.90     & 7244.40 \\
        \texttt{readelf}     & 29715.64 & \textbf{13012.36}       &  \textbf{29821.00} & 12990.70      & 29101.90 & 12813.30   & 29409.20 & 12928.80    & 29551.10   &  12934.20 \\
        \texttt{size}        & \textbf{3020.91} & \textbf{4030.91}         &  2984.20   & 3979.70      &  2999.60  & 3990.60   & 2975.20 & 4022.80     & 2925.60    & 4018.80\\
        \texttt{strip}       & \textbf{5732.55}  & 7703.55        &  5533.70  & 7667.60     & 5608.8  & \textbf{7716.20}        & 5542.80 &  7693.30    & 5549.30  & 7696.90\\
    \hline
    \end{tabular}
    \label{tab:eval_parameter}
\end{table*}

\paragraph{Orthogonality to Advanced Fuzzing Methods} 
To highlight \coolname's benefit in more recent fuzzers, we extend EcoFuzz~\cite{ecofuzz} with our mutation scheduler. EcoFuzz optimizes AFL's power schedule process to reduce AFL's focus on high-frequency paths. Within the fuzzer, we added four invocations to the DARWIN interface at the appropriate places in the code. We conducted a FuzzBench coverage experiment with 10 runs, 6h each. The full results are depicted in~\Cref{tab:eval_fuzzbench_ecofuzz}.
EcoFuzz-\coolname outperforms its baseline in all but four experiments. While \texttt{libjpeg-turbo-07-2017} and \texttt{systemd\_fuzz-link-parser} are quite close, the other two experiments show a larger difference.
The \coolname variant cannot outperform its baseline in \texttt{libxml2-v2.9.2} and \texttt{openthread-2019-12-23} \texttt{openssl\_x509}, similar as in the previous coverage experiment. Based on our investigation, this is also caused by the strongly structured input (openthread is an implementation of the OpenThread networking protocol), where the baseline fuzzer profits from higher execution speeds.

\begin{table}[!ht]
    \centering
    \caption{Median relative code coverage on each benchmark after 10 runs with 6h each. Median relative performance of each fuzzer to the encountered experiment maximum.}
    \begin{tabular}{|l|c|c|c|}
        \hline
        & \textbf{EcoFuzz-\coolname} & \textbf{EcoFuzz} \\ \hline
        \textbf{FuzzerMedian} & \textbf{97.31} & 95.43 \\ \hline
        \textbf{FuzzerMean} & \textbf{94.47} & 94.29 \\ \Xhline{1\arrayrulewidth}
        bloaty\_fuzz\_target & \textbf{95.86} & 91.39 \\ \hline
        curl\_curl\_fuzzer\_http & \textbf{97.31} & 95.89 \\ \hline
        freetype2-2017 & \textbf{95.92} & 95.79 \\ \hline
        harfbuzz-1.3.2 & \textbf{97.36} & 94.12 \\ \hline
        libjpeg-turbo-07-2017 & 84.67 & \textbf{86.47} \\ \hline
        libpng-1.2.56 & \textbf{97.83} & 96.51 \\ \hline
        libxml2-v2.9.2 & 78.84 & \textbf{93.66} \\ \hline
        libxslt\_xpath & \textbf{94.03} & 93.97 \\ \hline
        mbedtls\_fuzz\_dtlsclient & \textbf{99.23} & 96.71 \\ \hline
        openssl\_x509 & \textbf{99.67} & 99.62 \\ \hline
        openthread-2019-12-23 & 80.11 & \textbf{98.45} \\ \hline
        php\_php-fuzz-parser & \textbf{99.62} & 99.49 \\ \hline
        proj4-2017-08-14 & \textbf{90.53} & 84.77 \\ \hline
        re2-2014-12-09 & \textbf{98.83} & 98.50 \\ \hline
        sqlite3\_ossfuzz & \textbf{95.78} & 83.01 \\ \hline 
        systemd\_fuzz-link-parser & 97.96 & \textbf{98.90} \\ \hline
        vorbis-2017-12-11 & \textbf{96.09} & 95.43 \\ \hline
        woff2-2016-05-06 & \textbf{97.60} & 93.60 \\ \hline
        zlib\_zlib\_uncompress\_fuzzer & \textbf{97.70} & 95.24 \\ \hline
    \end{tabular}

    \label{tab:eval_fuzzbench_ecofuzz}
\end{table}

\subsection{Execution Speed versus Efficiency}
\label{subsec:eval_speed}

Challenge~\ref{itm:cperf} underlines the difficulty of optimizing probability distribution without spending too much time on a learning algorithm.
This is important as an optimal distribution does not lead to a measurable improvement if the optimal selection can be found via brute force in less time.
As such, we measure the effectiveness of the mutation scheduler in finding a good mutation probability distribution. Further, we analyze and compare the execution speed of \coolname's ES, MO{\small PT}\normalsize{}'s PSO, and AFL's random sampling with a uniform probability distribution.

\paragraph{Scheduling Effectiveness} While it is rather simple to measure the effects of an algorithmic change in fuzzing via coverage or crash analysis, the resulting numbers are hard to attribute to the algorithmic change itself due to the fuzzers complexity. Hence, we derived a metric to directly capture the impact of mutation scheduling, namely the average number of mutations needed to go from one coverage point to another.
Here, we get 1981.90 mutations for AFL, 1484.81 for MO{\small PT}\normalsize{}, and 1491.32 for \coolname. This clearly shows the advantage of mutation scheduling. MO{\small PT}\normalsize{} and \coolname achieve very similar results, where we attribute the difference to noise. 
The remaining question is whether both fuzzers also achieve the same execution speed, as the mutation schedulers' efficiency depends on both factors.

\begin{table}[!ht]
    \centering
    \caption{Averaged executions per second reached with the respective mutation scheduling approach.}
    \begin{tabular}{llrr}
    \hline
    Benchmark & havoc   & afl     & mopt    \\\hline
    bsdtar    & 2631.86 & 2385    & 1185.87 \\
    cxxfilt   & 2060.07 & 3766.8  & 2888.56 \\
    djpeg     & 2830.72 & 2609.9  & 1097.79 \\
    jhead     & 5097.98 & 5484.94 & 1679.58 \\
    objcopy   & 2019.37 & 2086.37 & 1867.22 \\
    objdump   & 1908.94 & 1932.47 & 1887.44 \\ \hline
    readelf   & 2439.79 & 2715.96 & 2389.04 \\
    size      & 2082.11 & 2147.08 & 1945.45 \\
    strip     & 2005.56 & 2159.54 & 1876.9  \\
    tcpdump   & 5042.22 & 5232.37 & 1554.85 \\ \hline
    geomean   &         & -7.6\%    & +48.26\%  \\ \hline
    \end{tabular}
    \label{tab:eval_speed}
\end{table}

\paragraph{Performance Measurements} \Cref{tab:eval_speed} presents the observed execution speed over ten runs.
Notably, AFL has the most executions, which makes sense considering that both \coolname and MO{\small PT}\normalsize{}  add an optimization algorithm on top of AFL's random sampling; yet, the \coolname's execution speed is relatively close to random selection.
However, the numbers demonstrate that \coolname is 48.26\% (geometric mean) faster than MO{\small PT}\normalsize{} while outperforming both other fuzzers in terms of coverage.
This makes \coolname solve Challenge~\ref{itm:cperf} and also highlights that the representation encoding for ES does not induce a major performance overhead.

This underlines that (1) \coolname's mutation scheduler improves efficiency compared to uniform random sampling and (2) that \coolname's mutation scheduler achieves this with less computational overhead than MO{\small PT}\normalsize, addressing Challenge~\ref{itm:algo}
In conclusion, \coolname has the same scheduling effectiveness but is much faster than MO{\small PT}\normalsize{}, resulting in better efficiency.

\subsection{MAGMA - Time-to-Bug Evaluation}
\label{subsec:eval_magma}
MAGMA~\cite{magma} is a recently published fuzzer benchmark that emphasizes the capability to uncover bugs, in particular, the time needed to reach a bug within a target.
For this, the authors forward-port real-world bugs into current versions of tools used in practice, namely \texttt{libpng}, \texttt{libtiff}, \texttt{libxml2}, \texttt{openssl} (which we could not get to run with the current version of MAGMA on GitHub at the time of writing, \texttt{php}, \texttt{poppler}, and \texttt{sqlite3}.
Further, MAGMA provides a framework around these tools to detect when a fuzzer reaches and triggers such a forward-ported bug.
Hence, MAGMA's attempt to measure the time to reach a bug gives a much clearer picture of a fuzzer's efficiency in practice, as code coverage is merely a proxy metric to measure a fuzzer's success.
We set up five hours fuzzing campaigns for each target for the MAGMA benchmark and repeated each experiment three times.

\begin{figure}[h!]
    \centering
    \includegraphics[width=1\linewidth]{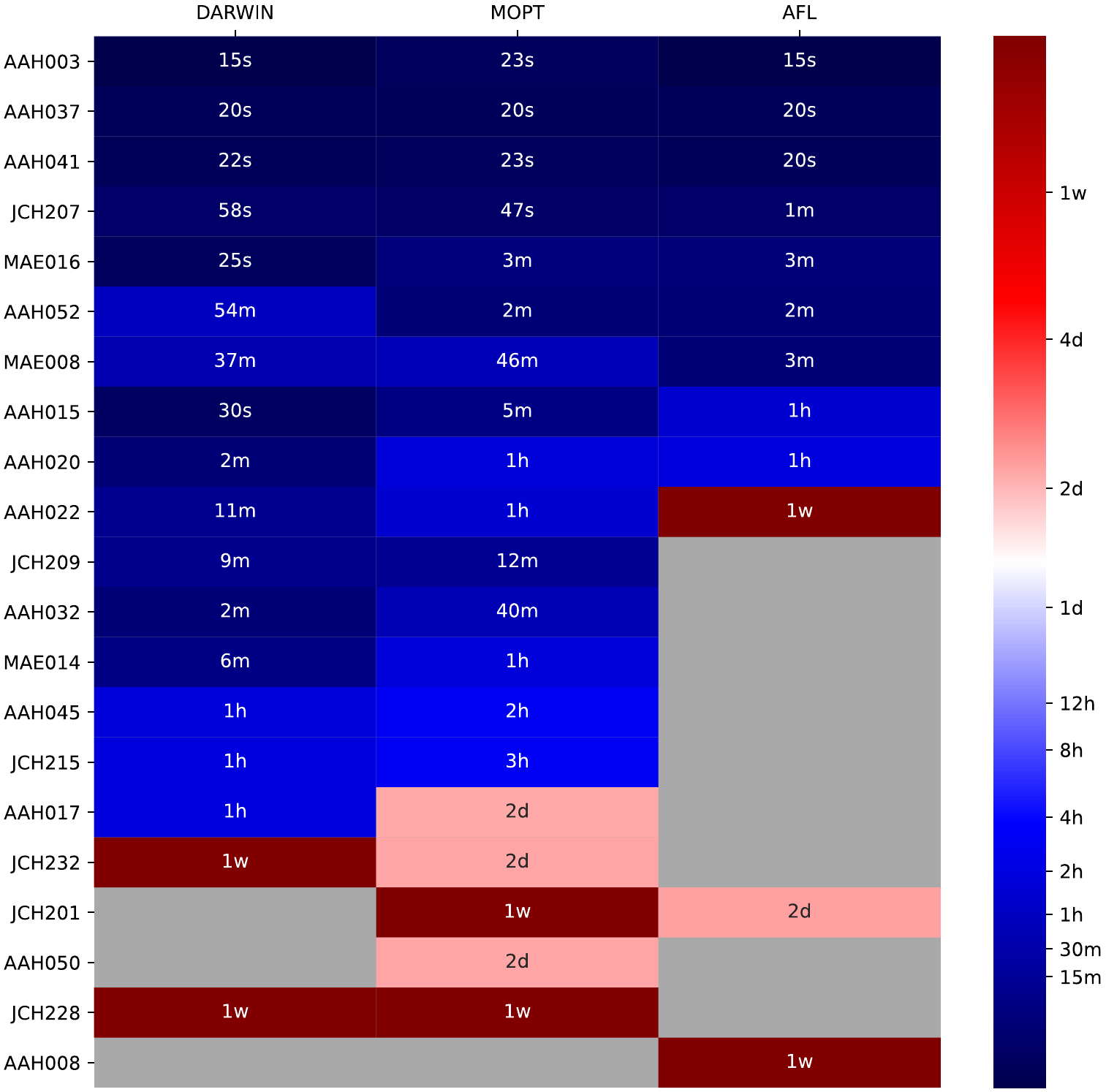}
    \caption{The expected time to reach a bug in the MAGMA benchmark over three runs. Y axis shows the individual bugs. Lower time is better, grey indicates that a fuzzer has not found this bug.}
    \label{fig:magma}
\end{figure}

The results are depicted in~\Cref{fig:magma}.
Out of 21 bugs found in total, \coolname can find 15 of them the fastest.
MO{\small PT}\normalsize{} is in 4 cases the fastest, but only because in two of them \coolname could not trigger the bug (where MO{\small PT}\normalsize{} is expected to take more than two days to find the bug on average).
Finally, AFL can only find 12 bugs, further emphasizing that \coolname increases the efficiency of the mutation selection.

\subsection{Crashes}
This final experiment explores \coolname's ability to find crashes in well-fuzzed targets, which is commonly done to evaluate fuzzers~\cite{ecofuzz,mopt,driller,vuzzer}. 
\changed{Note that our experiment differs from the setup MO{\small PT}\normalsize{} paper to increase statistical meaningfulness. In the MO{\small PT}\normalsize{} paper, the authors use 100 seed files per target. This, however, makes interpretation of the resulting data highly challenging, as the outcome heavily depends on which seed has been scheduled (also makes finding novel bugs much more likely). 
Additionally, the experiment was running only once.}
Here, We conducted a 24h experiment with 10 runs (and same seeds as in previous experiments) to also evaluate the stability in finding bugs. We use the same benchmarks as used in \Cref{subsec:cov} already.

The resulting crashes are shown in~\Cref{table:crashes}. Then, we minimized the test cases using afl-tmin and verified them with afl-collect~\cite{aflutils}. Then we first removed test cases with the same MD5 hash, and the Address Sanitizer output refers to the same line. Finally, we manually verified that they differ and lead to a crash, which we refer to as "triaged" in~\Cref{table:crashes}.
In total, we found 20 unique bugs with \coolname, and 26 unique bugs with the \coolname-enhanced version of EcoFuzz.
In contrast, the baselines, AFL and EcoFuzz, only found 12 resp. 1 unique bug(s). Also, the stability of their findings (i.e., the mean over all runs) is way below the \coolname-based fuzzers.\\

\changed{\coolname also found a completely novel bug in \texttt{objcopy} (working up to binutils 2.39, introduced more than 24 years ago), which is leading to a memory leak. \texttt{copy\_relocations\_in\_section} in objcopy.c is not freeing a buffer (relpp) in every possible case. This bug is very hard to trigger, as the function is only called at high stack depths. The testcase leading to the bug was found through splicing based on a relatively early testcase and a testcase from the middle of the experiment. We responsibly disclosed the triaged bug to the respective developers, who acknowledged and fixed the bug~\footnote{\url{https://sourceware.org/bugzilla/show_bug.cgi?id=29233}}.}

\begin{table*}[ht]
    \scriptsize
    \centering
    \caption{Crashes encountered in a 24h campaign over 10 runs. All bugs are triaged crashes. "Max" refers to the maximum encountered bugs within a run. "Uniq" refers to the number of unique crashes over all 10 runs. }
    \begin{tabular}{|l|r|r|r|r|r|r|r|r|r|r|r|r||r|r|r|r|r|r|r|}
        \hline
            &   \multicolumn{3}{c|}{\coolname} &    \multicolumn{3}{c|}{AFL} &    \multicolumn{3}{c|}{AFL-S} &    \multicolumn{3}{c||}{MOPT} &    \multicolumn{3}{c|}{EcoFuzz-D} &    \multicolumn{3}{c|}{EcoFuzz}\\
        \multicolumn{1}{|l|}{Benchmark} & \multicolumn{1}{c|}{mean} & \multicolumn{1}{c|}{max} & \multicolumn{1}{c|}{uniq} & \multicolumn{1}{c|}{mean} & \multicolumn{1}{c|}{max} & \multicolumn{1}{c|}{uniq}& \multicolumn{1}{c|}{mean} & \multicolumn{1}{c|}{max} & \multicolumn{1}{c|}{uniq}& \multicolumn{1}{c|}{mean} & \multicolumn{1}{c|}{max} & \multicolumn{1}{c||}{uniq} & \multicolumn{1}{c|}{mean} & \multicolumn{1}{c|}{max} & \multicolumn{1}{c|}{uniq} & \multicolumn{1}{c|}{mean} & \multicolumn{1}{c|}{max} & \multicolumn{1}{c|}{uniq} \\\hline
bsdtar                         & 0                          & 0                          & 0                     & 0                        & 0                        & 0                     & 0                          & 0                           & 0                     & 0                        & 0                         & 0                     & 0                                 & 0                                 & 0                    & 0                           & 0                           & 0                    \\
djpeg                          & 0                          & 0                          & 0                     & 0                        & 0                        & 0                     & 0                          & 0                           & 0                     & 0                        & 0                         & 0                     & 0                                 & 0                                 & 0                    & 0                           & 0                           & 0                    \\
tcpdump                        & 0                          & 0                          & 0                     & 0                        & 0                        & 0                     & 0                          & 0                           & 0                     & 0                        & 0                         & 0                     & 0                                 & 0                                 & 0                    & 0                           & 0                           & 0                    \\
jhead                          & 0                          & 0                          & 0                     & 0                        & 0                        & 0                     & 0                          & 0                           & 0                     & 0                        & 0                         & 0                     & 0                                 & 0                                 & 0                    & 0                           & 0                           & 0                    \\
readelf                        & 0                          & 0                          & 0                     & 0                        & 0                        & 0                     & 0                          & 0                           & 0                     & 0                        & 0                         & 0                     & 0                                 & 0                                 & 0                    & 0                           & 0                           & 0                    \\
strip                          & 0.25                       & \textbf{3}                 & \textbf{3}            & 0.09                     & 1                        & 1                     & \textbf{0.3}               & 2                           & \textbf{3}            & 0                        & 0                         & 0                     & 0                                 & 0                                 & 0                    & 0                           & 0                           & 0                    \\
size                           & \textbf{0.92}              & \textbf{2}                 & \textbf{11}           & 0.45                     & 1                        & 5                     & 0.7                        & 1                           & 7                     & 0.17                     & 1                         & 2                     & 0.3                               & 1                                 & 3                    & 0.1                         & 1                           & 1                    \\
filt                           & 0                          & 0                          & 0                     & 0                        & 0                        & 0                     & 0                          & 0                           & 0                     & 0                        & 0                         & 0                     & 0                                 & 0                                 & 0                    & 0                           & 0                           & 0                    \\
objdump                        & \textbf{0.25}              & \textbf{1}                 & \textbf{3}            & 0.09                     & 1                        & 3                     & 0.1                        & 1                           & 1                     & 0                        & 0                         & 0                     & 0.2                               & 1                                 & 2                    & 0                           & 0                           & 0                    \\
objcopy                        & \textbf{0.25}                       & 1                          & 3                     & 0.18                     & 1                        & 3                     & 0.1                        & 1                           & 1                     & 0                        & 0                         & 0                     & \textbf{2.1}                      & \textbf{17}                       & \textbf{21}          & 0                           & 0                           & 0                    \\ 
        \hline
        Total       &   & \textbf{7} & \textbf{20}    &  & 4 & 12 & & 5 & 12 & & 1 & 2 & & \textbf{18} & \textbf{26}  & & 1 & 1\\\hline
     \end{tabular}
   
    \label{table:crashes}
\end{table*}

\section{Related Work}
\label{sec:related}

Fuzzing is an active research domain but is also widely used in practice.
It has been improved in various areas, e.g., grammar-based fuzzing that also might use mutations~\cite{nautilus}, dedicated mutations, or program transformations for common roadblocks~\cite{redqueen,TaintScope,tfuzz}, or fuzzers for hard-to-fuzz software, e.g., due to hardware dependencies~\cite{firmwarefuzzing,firmwarefuzzing2}.
We consider these works orthogonal to ours.
Next, we restrict ourselves to the areas of mutation scheduling but also seed-selection algorithms, as the underlying approaches are often similar.
Finally, we compare \coolname to the presented works.

\subsection{Mutation Strategies}
Mutation strategies try to optimize either what mutations should be applied (which we refer to as \textit{mutation scheduling}) or where in the input those mutations should be applied (which we refer to as \textit{location optimization}).\\ %

\textbf{Mutation Scheduling.} In 2018, two works proposed to leverage machine learning approaches to improve mutation scheduling. B\"{o}ttinger et al. used deep Q-learning (a type of reinforcement learning) to find policies that can next generate new higher reward inputs~\cite{8424642}. Drozd and Wagner optimized mutation operators using reinforcement learning to achieve deeper coverage across several varied benchmarks~\cite{DBLP:journals/corr/abs-1807-07490}.
Despite leveraging complex algorithms, both of those works do not manage to show significant improvements in vulnerability discovery, underlining that the algorithms are too complex to address Challenges~\ref{itm:optimal} and~\ref{itm:cperf}.
Lyu et al. considered a different approach for optimizing mutation scheduling and proposed a mutation scheduling scheme called MO\small{PT}~\cite{mopt}. MO\small{PT} \normalsize was the first work to propose using heuristic techniques for optimizing general mutation scheduling. More precisely, 
MO\small{PT} \normalsize uses a custom variant of Particle Swarm Optimization (PSO) to approximate the best selection probability distribution for mutation operators.
We note that for PSO, there is no guarantee to converge to the global optimum (only to the best particle in the swarm)~\cite{10.5555/935867,DBLP:books/daglib/0024369}.
At the same time, there are proofs of convergence for evolution strategy~\cite{Hansen2015}. 
Further, MO\small{PT}  \normalsize proposes to deactivate the deterministic fuzzing stage either temporarily or permanently to make PSO converge faster.

As this work is closest in the objective and applied techniques to ours, we discuss the main differences between MO\small{PT}  \normalsize and \coolname in more detail .
From MO\small{PT}\normalsize 's design perspective, the authors do not show how several parameters need to be tuned to reach a good performance under which condition. 
In particular, it is not evaluated how many solutions (swarms) are needed in practice and how difficult it is to tune them, or how sensitive those parameters are. Hence, MO\small{PT}\normalsize{} does not solve Challenge~\ref{itm:adoption}. %
Since the MO\small{PT}  \normalsize algorithm has both local and global positions for particles, the algorithm requires additional measures to find the best solutions, increasing the complexity of the algorithm.
This leads to a performance reduction in the havoc stage, as we explore in~\Cref{subsec:eval_speed}. Thus, MO\small{PT}\normalsize{} cannot address Challenge~\ref{itm:cperf}.
A change of solution encoding, as proposed in \Cref{subsec:encoding}, requires changes in MO\small{PT}  \normalsize's  algorithm.
Finally, what the authors call a swarm is actually a solution in a swarm. What the authors denote as multiple swarms is one swarm.

In contrast, \coolname has no parameters to tune from the fuzzer side. ES has only two parameters, $\mu$ and $\lambda$, which are intuitive to select during fuzzer development time and have a clear role in the evolution process. \coolname does not require any additional communication between modules to run the evolution process. \coolname uses a simple fitness function where the goal is the maximization of the code coverage. \coolname supports various solution representations without requiring changes in the \coolname algorithm. We develop \coolname not only to be well-performing for the specific application at hand but also to conform to standards from the EA community regarding the design choices and performance evaluation.

From the performance perspective, MO{\small PT}\normalsize's PSO integration is computationally intense (i.e., already reducing coverage significantly over time due to decreased speed), and the evaluation does not explore whether a simpler algorithm or even a static distribution might already be enough.
The evaluation results are also produced by a varying amount of seed files, but not a typical setup with one empty and one small seed suitable for the application.
Further, MO{\small PT}\normalsize's mutation scheduling algorithm is not evaluated separately from the other stages of the fuzzer but always with the deterministic stage running at least once.

Our evaluation shows that these two aspects distorted the comparison with the default random mutation selection by microbenchmarking the mutation selection using our proposed average-mutations to a new coverage metric.
Besides, the huge size of seeds might lead to a distortion in coverage measurements since a fuzzer might be stuck for a while given a bad randomly chosen seed.
While we consider MO{\small PT}\normalsize's pacemaker mode as orthogonal, we still show that with a permanently disabled deterministic stage, AFL discovers significantly more unique paths than MO{\small PT}\normalsize, which is in line with the results reported in Google's FuzzBench~\cite{fuzzbench-report}.
In contrast, \coolname's selection algorithm is much simpler, has, thanks to its more lightweight Evolution Strategy and solution representation, a lower impact on execution speed, brings a measurable improvement over the standard uniform mutation selection, and even outperforms MO{\small PT}\normalsize{} significantly in terms of coverage and crashes found.\\

\textbf{Location Optimization.} In contrast to mutation scheduling approaches, some works aim to find the right locations in the inputs to mutate.
One example is FairFuzz which applies a deterministic combination of mutations to explore which bytes in the test case reach rare branches when mutated~\cite{fairfuzz}.
These bytes now form a mask used in the havoc stage to (partially) limit mutation operators to these bytes.
A similar approach has been proposed by Rajpal et al.~\cite{rajpal2017not}, where neural networks are used to infer (un-)promising bytes in inputs generated by past mutations. 
Promising bytes are then preferred during mutation.
Another work, Steelix~\cite{steelix}, leverages static analysis to extract information about comparisons in the target program, which is then used to mutate responsible bytes in the input efficiently.
Analogous to FairFuzz, the information generated by the static analysis is used to create a mask.
If a mutated input does not generate new coverage, but a byte in the mask is closer to what the comparison expects, this byte is further mutated.
All of these approaches above focus on where to apply mutations, whereas \coolname optimizes general mutation selection.
Further, many of the mentioned ideas can be combined with our approach.

\subsection{Seed-selection Algorithms}
Seed-selection algorithms aim to distill and select a subset of seeds to optimize for a specific branch to pass or improve coverage in general by preferring more promising seeds or minimizing seeds to improve execution speed.
MoonShine uses system call traces of real-world programs to distill them into a minimal test case that still achieves 86\% of the pre-distilled coverage~\cite{moonshine}.
These minimal tests can then be used to 1) trigger basic blocks that require a certain order of system calls and 2) improve the fuzzing speed.

A similar idea is used by FasterFuzzing, which employs a Generative Adversarial Network trained with an initial seed corpus to generate new, better seeds~\cite{fasterfuzzing}.
EcoFuzz~\cite{ecofuzz} proposes a seed scheduling algorithm to fine-tune exploration and exploitation.
After a short fuzzing period, EcoFuzz switches to the exploration phase, where the remaining seeds are fuzzed to estimate their reward probability.
Then, EcoFuzz switches to the exploitation phase to fuzz these seeds that have the highest reward probability.
If a new path has been discovered, EcoFuzz switches back to the exploration phase.
This increases coverage while reducing the number of test case generations.
AFLFast identifies that fuzzers are often stuck with high-frequency paths~\cite{aflfast}.
To balance this, AFLFast leverages a Markov model to identify and prefer low-frequency paths as a heuristic. 
Similarly, VUzzer uses an evolutionary algorithm approach to leverage control-flow features and find hard-to-reach paths while also avoiding inputs that reach basic blocks containing error-handling code~\cite{vuzzer}.
NeuFuzz, instead, does not try to balance low- and high-frequency paths but uses a neural network to prefer paths that are prone to contain vulnerabilities~\cite{neufuzz}.
Angora follows a more general strategy by preferring inputs that lead to unexplored branches, effectively also balancing high- and low-frequency path exploration~\cite{angora}.
AFLSmart uses a structural representation of seed to perform semantically correct mutations and increases time spent on mutating promising seeds that pass the input parsing~\cite{smartgreybox}.
AFLGo~\cite{aflgo} enables directed fuzzing close to chosen target locations by prioritizing seeds that reach paths close to the target~\cite{aflgo}. 
Seed-scheduling and -distilling algorithms optimize an early stage in the fuzzing process. Hence, it is challenging for these techniques to steer the mutation phase unless the havoc stage is specifically aware of, e.g., the phases defined in EcoFuzz.
This might lead to counterproductive mutations being applied to optimized seeds, canceling out the desired effect.
In contrast, \coolname optimizes a late stage in the fuzzing process and thus, can learn a favorable probability distribution to keep the properties of promising inputs.

\subsection{Algorithmic Improvements vs. Optimizing Execution Speed}
\label{subsec:rw_algo_improv}
Many works recently focused on the raw speed of input generation and mutation with big coverage improvements~\cite{kAFL,vecemu,redqueen,gopinath2019building}.
While \coolname offers fewer coverage improvements as reported by these fuzzers, \coolname's mutation scheduling is orthogonal to performance increases achieved through, e.g., fast snapshotting.
Hence, \coolname can further increase coverage, and more importantly---as we show in~\Cref{subsec:eval_lava} and~\Cref{subsec:eval_magma}---improve the bug triggering capabilities of these fuzzers.

\section{Conclusion}

We presented \coolname, a novel mutation scheduling algorithm that uses an Evolution Strategy to optimize the mutation selection probability distribution based on the instrumented application's feedback. \coolname tackles all of our identified challenges in building a mutation scheduler: Challenge~\ref{itm:optimal} by integrating Evolutionary Strategy as a mutation scheduler, significantly outperforming the state-of-the-art mutation scheduler MO{\small PT}\normalsize{}~\cite{mopt}, while also being the first mutation scheduler to show a significant increase in edge coverage of 1.73\% over AFL respectively, bugs uncovered in both, LAVA-M and MAGMA, and decrease in time to find bugs over AFL and MO{\small PT}\normalsize{}; Challenge~\ref{itm:algo} by choosing reasonable encoding and parameters; Challenge~\ref{itm:adoption} by introducing no user-facing parameters that need to be tuned per target; and Challenge~\ref{itm:cperf} by maintaining a high execution speed compared to the AFL baseline, in contrast to related work, which is far slower.
Further, \coolname found 20 unique bugs in widely-used real-world applications, outperforming both AFL and MO{\small PT}\normalsize{}.
\coolname was the only fuzzer able to also uncover a new bug that is still working on the most recent version of the target.
While our experiments show that unique path coverage for fitness provides good feedback for ES, other heuristics could be used. For example, it would be interesting to include the number of crashes and consider the Pareto fronts of the solutions.
Further, future research could study the efficiency of multi-objective algorithms for mutation scheduling that combine several of the previous suggestions, e.g., also include the frequency of the path to focus more on low-frequency paths or the block hit count to promote stronger intensification. 
\section*{Acknowledgments}
This work was supported by the German Federal Ministry of Education and Research in the StartUpSecure funding program "Sanctuary" (16KIS1417), the German Federal Ministry of Education and Research and the Hessian State Ministry for Higher Education, Research and the Arts within ATHENE, and by the European Research Council (ERC) under the European Union's Horizon 2020 research and innovation programme (grant agreement No.\ 952697).

\bibliographystyle{plain}
\bibliography{bib}

\begin{thebibliography}{10}

\bibitem{Abouhawwash572}
Mohamed Abouhawwash, Kalyanmoy Deb, and Adam Alessio.
\newblock Exploration of multi-objective optimization with genetic algorithms
  for pet image reconstruction.
\newblock {\em Journal of Nuclear Medicine}, 61(supplement 1):572--572, 2020.

\bibitem{arcuri2014hitchhiker}
Andrea Arcuri and Lionel Briand.
\newblock A hitchhiker's guide to statistical tests for assessing randomized
  algorithms in software engineering.
\newblock {\em Software Testing, Verification and Reliability}, 24(3):219--250,
  2014.

\bibitem{nautilus}
Cornelius Aschermann, Tommaso Frassetto, Thorsten Holz, Patrick Jauernig,
  Ahmad-Reza Sadeghi, and Daniel Teuchert.
\newblock Nautilus: Fishing for deep bugs with grammars.
\newblock In {\em NDSS}, 2019.

\bibitem{redqueen}
Cornelius Aschermann, Sergej Schumilo, Tim Blazytko, Robert Gawlik, and
  Thorsten Holz.
\newblock Redqueen: Fuzzing with input-to-state correspondence.
\newblock In {\em NDSS}, volume~19, pages 1--15, 2019.

\bibitem{1688465}
H.-G. Beyer and B.~Sendhoff.
\newblock Evolution strategies for robust optimization.
\newblock In {\em 2006 IEEE International Conference on Evolutionary
  Computation}, pages 1346--1353, 2006.

\bibitem{10.1023/A:1015059928466}
Hans-Georg Beyer and Hans-Paul Schwefel.
\newblock {\em Evolution Strategies – A Comprehensive Introduction},
  volume~1.
\newblock Kluwer Academic Publishers, USA, May 2002.

\bibitem{aflgo}
Marcel B{\"o}hme, Van-Thuan Pham, Manh-Dung Nguyen, and Abhik Roychoudhury.
\newblock Directed greybox fuzzing.
\newblock In {\em Proceedings of the 2017 ACM SIGSAC Conference on Computer and
  Communications Security}, pages 2329--2344, 2017.

\bibitem{aflfast}
Marcel B{\"o}hme, Van-Thuan Pham, and Abhik Roychoudhury.
\newblock Coverage-based greybox fuzzing as markov chain.
\newblock In {\em ACM Conference on Computer and Communications Security
  (CCS)}, 2016.

\bibitem{7101236}
Jürgen Branke, Su~Nguyen, Christoph~W. Pickardt, and Mengjie Zhang.
\newblock Automated design of production scheduling heuristics: A review.
\newblock {\em IEEE Transactions on Evolutionary Computation}, 20(1):110--124,
  2016.

\bibitem{8424642}
K.~{Böttinger}, P.~{Godefroid}, and R.~{Singh}.
\newblock Deep reinforcement fuzzing.
\newblock In {\em 2018 IEEE Security and Privacy Workshops (SPW)}, pages
  116--122, 2018.

\bibitem{hawkeye}
Hongxu Chen, Yinxing Xue, Yuekang Li, Bihuan Chen, Xiaofei Xie, Xiuheng Wu, and
  Yang Liu.
\newblock Hawkeye: Towards a desired directed grey-box fuzzer.
\newblock In {\em Proceedings of the 2018 ACM SIGSAC Conference on Computer and
  Communications Security}, pages 2095--2108, 2018.

\bibitem{angora}
Peng Chen and Hao Chen.
\newblock Angora: Efficient fuzzing by principled search.
\newblock In {\em IEEE Symposium on Security and Privacy}, 2018.

\bibitem{chen2019ptrix}
Yaohui Chen, Dongliang Mu, Jun Xu, Zhichuang Sun, Wenbo Shen, Xinyu Xing, Long
  Lu, and Bing Mao.
\newblock Ptrix: Efficient hardware-assisted fuzzing for cots binary.
\newblock In {\em Proceedings of the 2019 ACM Asia Conference on Computer and
  Communications Security}, pages 633--645, 2019.

\bibitem{dolan2016lava}
Brendan Dolan-Gavitt, Patrick Hulin, Engin Kirda, Tim Leek, Andrea Mambretti,
  Wil Robertson, Frederick Ulrich, and Ryan Whelan.
\newblock Lava: Large-scale automated vulnerability addition.
\newblock In {\em 2016 IEEE Symposium on Security and Privacy (SP)}, pages
  110--121. IEEE, 2016.

\bibitem{DBLP:journals/corr/abs-1807-07490}
William Drozd and Michael~D. Wagner.
\newblock Fuzzergym: {A} competitive framework for fuzzing and learning.
\newblock {\em CoRR}, abs/1807.07490, 2018.

\bibitem{Eiben03}
A.~E. Eiben and J.~E. Smith.
\newblock {\em {Introduction to Evolutionary Computing}}.
\newblock Springer-Verlag, Berlin Heidelberg New York, USA, 2003.

\bibitem{emmerich2018evolution}
Michael Emmerich, Ofer~M Shir, and Hao Wang.
\newblock {\em Evolution Strategies}, chapter~4, pages 1--31.
\newblock Springer International Publishing, 2018.

\bibitem{DBLP:books/daglib/0024369}
Andries~P. Engelbrecht.
\newblock {\em Fundamentals of Computational Swarm Intelligence}.
\newblock Wiley, 2005.

\bibitem{vecemu}
Brandon Falk.
\newblock Vectorized emulation: Hardware accelerated taint tracking at 2
  trillion instructions per second.
\newblock
  \url{https://gamozolabs.github.io/fuzzing/2018/10/14/vectorized_emulation.html}.
\newblock Accessed: 2022-04-26.

\bibitem{firmwarefuzzing2}
Bo~Feng, Alejandro Mera, and Long Lu.
\newblock P 2 im: Scalable and hardware-independent firmware testing via
  automatic peripheral interface modeling.
\newblock In {\em Proceedings of the 29th USENIX Security Symposium}, 2020.

\bibitem{afl++}
Andrea Fioraldi, Dominik Maier, Heiko Ei{\ss}feldt, and Marc Heuse.
\newblock Afl++: Combining incremental steps of fuzzing research.
\newblock In {\em 14th USENIX Workshop on Offensive Technologies (WOOT 20)},
  2020.

\bibitem{doi:10.1080/0305215X.2019.1651310}
Abhinav Gaur, A.K.M.~Khaled Talukder, Kalyanmoy Deb, Santosh Tiwari, Simon Xu,
  and Don Jones.
\newblock Unconventional optimization for achieving well-informed design
  solutions for the automobile industry.
\newblock {\em Engineering Optimization}, 52(9):1542--1560, 2020.

\bibitem{10.1080/23311916.2014.945820}
Morteza Gholamipoor, Parviz Ghadimi, Mohammad~H. Alavidoost, and Mohammad
  A.~Feizi Chekab.
\newblock Application of evolution strategy algorithm for optimization of a
  single-layer sound absorber.
\newblock {\em Cogent Engineering}, 1(1):945820, 2014.

\bibitem{9185861}
Abhiroop Ghosh, Erik Goodman, Kalyanmoy Deb, Ronald Averill, and Alejandro
  Diaz.
\newblock A large-scale bi-objective optimization of solid rocket motors using
  innovization.
\newblock In {\em 2020 IEEE Congress on Evolutionary Computation (CEC)}, pages
  1--8, 2020.

\bibitem{10.5555/549765}
Fred Glover and Manuel Laguna.
\newblock {\em Tabu Search}.
\newblock Kluwer Academic Publishers, USA, 1997.

\bibitem{meta}
Fred~W. Glover and Gary~A. Kochenberger, editors.
\newblock {\em {Handbook of Metaheuristics}}, volume 114 of {\em International
  Series in Operations Research \& Management Science}.
\newblock Springer, 1 edition, January 2003.

\bibitem{fuzzbench-report}
Google.
\newblock Fuzzbench: 2020-09-28 report.
\newblock \url{https://www.fuzzbench.com/reports/2022-04-19/index.html}.
\newblock Accessed: 2022-04-26.

\bibitem{google-ossfuzz}
Google.
\newblock Oss-fuzz.
\newblock \url{https://google.github.io/oss-fuzz/}.
\newblock Accessed: 2022-04-26.

\bibitem{afl}
Google.
\newblock {\em american fuzzy loop (afl)}.
\newblock https://github.com/google/AFL, 2020.

\bibitem{gopinath2019building}
Rahul Gopinath and Andreas Zeller.
\newblock Building fast fuzzers.
\newblock {\em arXiv preprint arXiv:1911.07707}, 2019.

\bibitem{Hansen2015}
Nikolaus Hansen, Dirk~V. Arnold, and Anne Auger.
\newblock {\em Evolution Strategies}, pages 871--898.
\newblock Springer Berlin Heidelberg, Berlin, Heidelberg, 2015.

\bibitem{magma}
Ahmad Hazimeh, Adrian Herrera, and Mathias Payer.
\newblock Magma: A ground-truth fuzzing benchmark.
\newblock {\em Proceedings of the ACM on Measurement and Analysis of Computing
  Systems}, 4(3):1--29, 2020.

\bibitem{10.1080/03052150500035658}
Xiaolin Hu, Carlos A~Coello Coello, and Zhangcan Huang.
\newblock A new multi-objective evolutionary algorithm: Neighbourhood exploring
  evolution strategy.
\newblock {\em Engineering Optimization}, 37(4):351--379, 2005.

\bibitem{klees2018evaluating}
George Klees, Andrew Ruef, Benji Cooper, Shiyi Wei, and Michael Hicks.
\newblock Evaluating fuzz testing.
\newblock In {\em Proceedings of the 2018 ACM SIGSAC Conference on Computer and
  Communications Security}, 2018.

\bibitem{10.1007/978-3-030-16692-2_15}
Walter Krawec, Stjepan Picek, and Domagoj Jakobovic.
\newblock Evolutionary algorithms for the design of quantum protocols.
\newblock In Paul Kaufmann and Pedro~A. Castillo, editors, {\em Applications of
  Evolutionary Computation}, pages 220--236, Cham, 2019. Springer International
  Publishing.

\bibitem{10.5555/59580}
P.~J.~M. Laarhoven and E.~H.~L. Aarts.
\newblock {\em Simulated Annealing: Theory and Applications}.
\newblock Kluwer Academic Publishers, USA, 1987.

\bibitem{fairfuzz}
Caroline Lemieux and Koushik Sen.
\newblock Fairfuzz: A targeted mutation strategy for increasing greybox fuzz
  testing coverage.
\newblock In {\em Proceedings of the 33rd ACM/IEEE International Conference on
  Automated Software Engineering}, pages 475--485, 2018.

\bibitem{steelix}
Yuekang Li, Bihuan Chen, Mahinthan Chandramohan, Shang-Wei Lin, Yang Liu, and
  Alwen Tiu.
\newblock Steelix: program-state based binary fuzzing.
\newblock In {\em Proceedings of the 2017 11th Joint Meeting on Foundations of
  Software Engineering}, pages 627--637, 2017.

\bibitem{mopt}
Chenyang Lyu, Shouling Ji, Chao Zhang, Yuwei Li, Wei-Han Lee, Yu~Song, and
  Raheem Beyah.
\newblock {MOPT}: Optimized mutation scheduling for fuzzers.
\newblock In {\em 28th {USENIX} Security Symposium ({USENIX} Security 19)},
  pages 1949--1966, 2019.

\bibitem{fuzzbench}
Jonathan Metzman, L{\'a}szl{\'o} Szekeres, Laurent Simon, Read Sprabery, and
  Abhishek Arya.
\newblock Fuzzbench: an open fuzzer benchmarking platform and service.
\newblock In {\em Proceedings of the 29th ACM Joint Meeting on European
  Software Engineering Conference and Symposium on the Foundations of Software
  Engineering}, pages 1393--1403, 2021.

\bibitem{ms-onefuzz}
Microsoft.
\newblock Microsoft announces new project onefuzz framework, an open source
  developer tool to find and fix bugs at scale.
\newblock
  \url{https://www.microsoft.com/security/blog/2020/09/15/microsoft-onefuzz-framework-open-source-developer-tool-fix-bugs/}.
\newblock Accessed: 2022-04-26.

\bibitem{Miller2011}
Julian~F. Miller.
\newblock {\em Cartesian Genetic Programming}, pages 17--34.
\newblock Springer Berlin Heidelberg, Berlin, Heidelberg, 2011.

\bibitem{10.5555/522098}
Melanie Mitchell.
\newblock {\em An Introduction to Genetic Algorithms}.
\newblock MIT Press, Cambridge, MA, USA, 1998.

\bibitem{fasterfuzzing}
Nicole Nichols, Mark Raugas, Robert Jasper, and Nathan Hilliard.
\newblock Faster fuzzing: Reinitialization with deep neural models.
\newblock {\em arXiv preprint arXiv:1711.02807}, 2017.

\bibitem{convergence}
Beatrice Ombuki-Berman and Franklin Hanshar.
\newblock {\em Using Genetic Algorithms for Multi-depot Vehicle Routing},
  volume 161, pages 77--99.
\newblock Springer Berlin Heidelberg, 09 2008.

\bibitem{overton2020romu}
Mark~A Overton.
\newblock Romu: Fast nonlinear pseudo-random number generators providing high
  quality.
\newblock {\em arXiv preprint arXiv:2002.11331}, 2020.

\bibitem{owasp}
Inc. OWASP~Foundation.
\newblock Owasp top ten 2017.
\newblock
  \url{https://owasp.org/www-project-top-ten/2017/A9_2017-Using_Components_with_Known_Vulnerabilities}.
\newblock Accessed: 2022-04-26.

\bibitem{moonshine}
Shankara Pailoor, Andrew Aday, and Suman Jana.
\newblock {Moonshine: Optimizing OS fuzzer seed selection with trace
  distillation}.
\newblock In {\em 27th USENIX Security Symposium (USENIX Security 18)}, pages
  729--743, 2018.

\bibitem{tfuzz}
Hui Peng, Yan Shoshitaishvili, and Mathias Payer.
\newblock T-fuzz: fuzzing by program transformation.
\newblock In {\em 2018 IEEE Symposium on Security and Privacy (SP)}, pages
  697--710. IEEE, 2018.

\bibitem{smartgreybox}
Van-Thuan Pham, Marcel B{\"o}hme, Andrew~Edward Santosa, Alexandru~Razvan
  Caciulescu, and Abhik Roychoudhury.
\newblock Smart greybox fuzzing.
\newblock {\em IEEE Transactions on Software Engineering}, 2019.

\bibitem{rajpal2017not}
Mohit Rajpal, William Blum, and Rishabh Singh.
\newblock Not all bytes are equal: Neural byte sieve for fuzzing.
\newblock {\em arXiv preprint arXiv:1711.04596}, 2017.

\bibitem{vuzzer}
Sanjay Rawat, Vivek Jain, Ashish Kumar, Lucian Cojocar, Cristiano Giuffrida,
  and Herbert Bos.
\newblock Vuzzer: Application-aware evolutionary fuzzing.
\newblock In {\em Proceedings of the Network and Distributed System Security
  Symposium (NDSS)}, 2017.

\bibitem{aflutils}
rc0r.
\newblock afl-utils.
\newblock \url{https://gitlab.com/rc0r/afl-utils/-/tree/master/afl_utils}.
\newblock Accessed: 2022-04-26.

\bibitem{10.1007/978-3-030-58115-2_4}
Lino Rodriguez-Coayahuitl, Alicia Morales-Reyes, Hugo~Jair Escalante, and
  Carlos~A. Coello~Coello.
\newblock Cooperative co-evolutionary genetic programming for high dimensional
  problems.
\newblock In Thomas B{\"a}ck, Mike Preuss, Andr{\'e} Deutz, Hao Wang, Carola
  Doerr, Michael Emmerich, and Heike Trautmann, editors, {\em Parallel Problem
  Solving from Nature -- PPSN XVI}, pages 48--62, Cham, 2020. Springer
  International Publishing.

\bibitem{Kennedy2010}
Claude Sammut and Geoffrey~I. Webb, editors.
\newblock {\em Particle Swarm Optimization}.
\newblock Springer US, Boston, MA, 2010.

\bibitem{kAFL}
Sergej Schumilo, Cornelius Aschermann, Robert Gawlik, Sebastian Schinzel, and
  Thorsten Holz.
\newblock {kafl: Hardware-assisted feedback fuzzing for OS kernels}.
\newblock In {\em 26th USENIX Security Symposium (USENIX Security 17)}, pages
  167--182, 2017.

\bibitem{libfuzzer}
Kosta Serebryany.
\newblock Continuous fuzzing with libfuzzer and addresssanitizer.
\newblock In {\em 2016 IEEE Cybersecurity Development (SecDev)}, pages
  157--157. IEEE, 2016.

\bibitem{driller}
Nick Stephens, John Grosen, Christopher Salls, Andrew Dutcher, Ruoyu Wang,
  Jacopo Corbetta, Yan Shoshitaishvili, Christopher Kruegel, and Giovanni
  Vigna.
\newblock Driller: Augmenting fuzzing through selective symbolic execution.
\newblock In {\em NDSS}, volume~16, pages 1--16, 2016.

\bibitem{8601309}
Jiawei Su, Danilo~Vasconcellos Vargas, and Kouichi Sakurai.
\newblock One pixel attack for fooling deep neural networks.
\newblock {\em IEEE Transactions on Evolutionary Computation}, 23(5):828--841,
  2019.

\bibitem{Talbi}
El-Ghazali Talbi.
\newblock {\em Metaheuristics: From Design to Implementation}.
\newblock Wiley Publishing, 2009.

\bibitem{10.5555/935867}
Frans Van Den~Bergh and A.~P. Engelbrecht.
\newblock {\em An Analysis of Particle Swarm Optimizers}.
\newblock PhD thesis, ZAF, 2002.
\newblock AAI0804353.

\bibitem{syzkaller}
Dmitry Vyukov.
\newblock syzkaller - kernel fuzzer.
\newblock \url{https://github.com/google/syzkaller}.
\newblock Accessed: 2022-04-26.

\bibitem{TaintScope}
Tielei Wang, Tao Wei, Guofei Gu, and Wei Zou.
\newblock Taintscope: A checksum-aware directed fuzzing tool for automatic
  software vulnerability detection.
\newblock In {\em Security and privacy (SP), 2010 IEEE symposium on}, pages
  497--512. IEEE, 2010.

\bibitem{neufuzz}
Yunchao Wang, Zehui Wu, Qiang Wei, and Qingxian Wang.
\newblock Neufuzz: Efficient fuzzing with deep neural network.
\newblock {\em IEEE Access}, 7:36340--36352, 2019.

\bibitem{10.1007/978-3-030-40186-3_8}
Lichao Wu, Gerard Ribera, Noemie Beringuier-Boher, and Stjepan Picek.
\newblock A fast characterization method for semi-invasive fault injection
  attacks.
\newblock In Stanislaw Jarecki, editor, {\em Topics in Cryptology -- CT-RSA
  2020}, pages 146--170, Cham, 2020. Springer International Publishing.

\bibitem{wu2022one}
Mingyuan Wu, Ling Jiang, Jiahong Xiang, Yanwei Huang, Heming Cui, Lingming
  Zhang, and Yuqun Zhang.
\newblock One fuzzing strategy to rule them all.
\newblock In {\em Proceedings of the International Conference on Software
  Engineering}, 2022.

\bibitem{ecofuzz}
Tai Yue, Pengfei Wang, Yong Tang, Enze Wang, Bo~Yu, Kai Lu, and Xu~Zhou.
\newblock {EcoFuzz: Adaptive Energy-Saving Greybox Fuzzing as a Variant of the
  Adversarial Multi-Armed Bandit}.
\newblock In {\em 29th USENIX Security Symposium (USENIX Security 20)}, 2020.

\bibitem{qsym}
Insu Yun, Sangho Lee, Meng Xu, Yeongjin Jang, and Taesoo Kim.
\newblock {QSYM: A practical concolic execution engine tailored for hybrid
  fuzzing}.
\newblock In {\em 27th USENIX Security Symposium (USENIX Security 18)}, pages
  745--761, 2018.

\bibitem{firmwarefuzzing}
Yaowen Zheng, Ali Davanian, Heng Yin, Chengyu Song, Hongsong Zhu, and Limin
  Sun.
\newblock {FIRM-AFL: high-throughput greybox fuzzing of iot firmware via
  augmented process emulation}.
\newblock In {\em 28th USENIX Security Symposium (USENIX Security 19)}, pages
  1099--1114, 2019.

\bibitem{fuzzguard}
Peiyuan Zong, Tao Lv, Dawei Wang, Zizhuang Deng, Ruigang Liang, and Kai Chen.
\newblock Fuzzguard: Filtering out unreachable inputs in directed grey-box
  fuzzing through deep learning.
\newblock In {\em 29th {USENIX} Security Symposium ({USENIX} Security 20)},
  pages 2255--2269. {USENIX} Association, August 2020.

\end{thebibliography}

\appendix
\subsection{Evolutionary Algorithms}
\label{app:ea}

\begin{figure}[h]
    \centering
    \subfigure[\texttt{Depiction of perturbation for real-valued vector. The sum of all values does not need to be equal to 1 and every gene must have a non-negative value. }]{\includegraphics[width=0.41\textwidth]{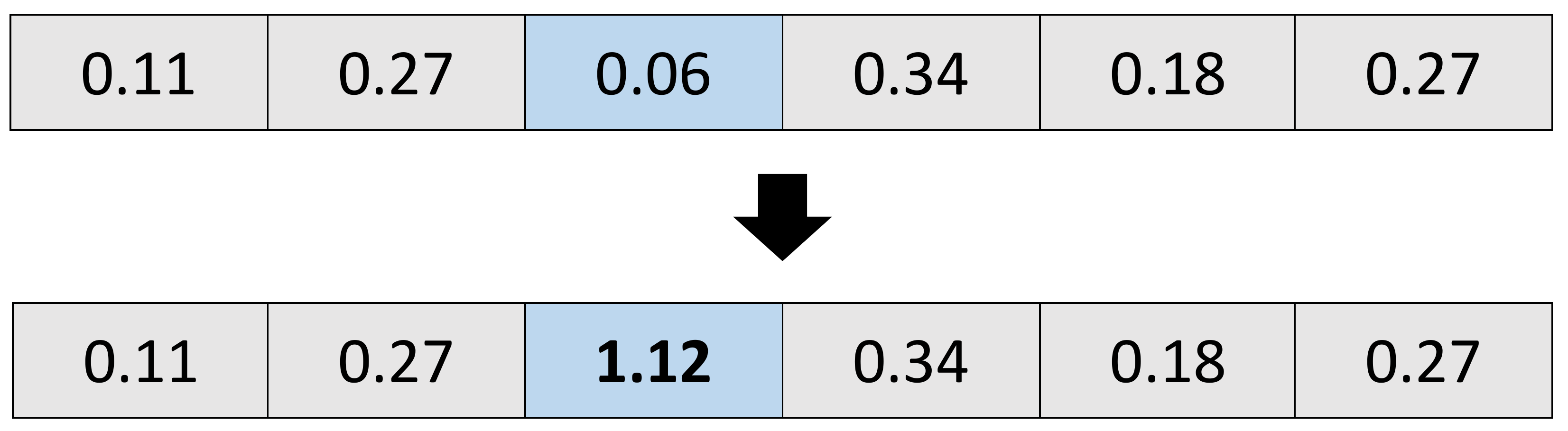}\label{fig:float_encoding}} 
    \subfigure[\texttt{Depiction of perturbation for binary vector.}]{\includegraphics[width=0.41\textwidth]{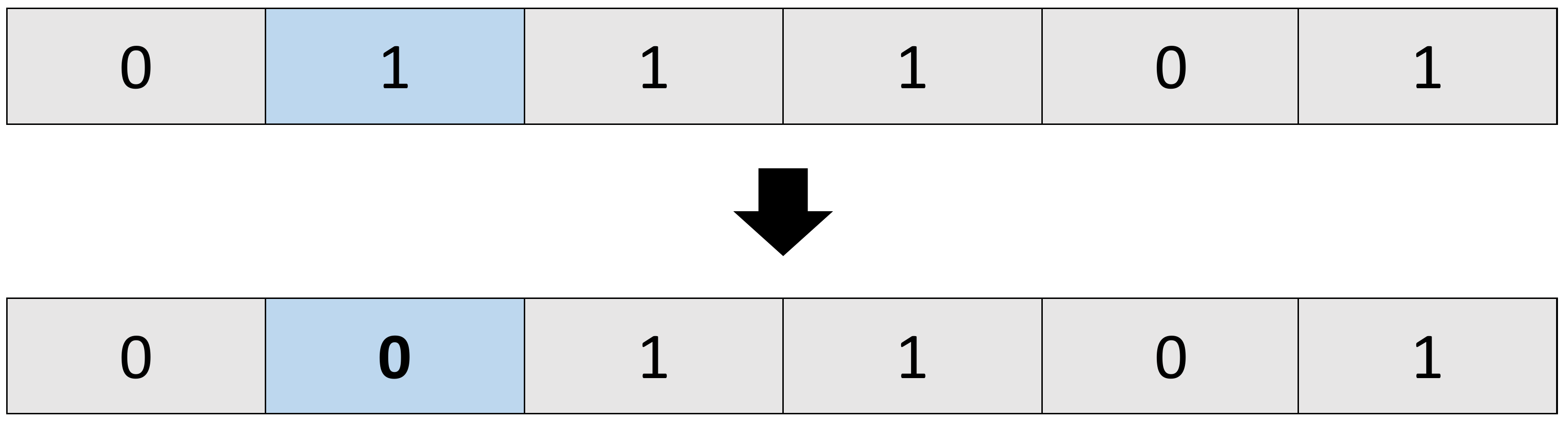}\label{fig:binary_encoding}}
    \caption{Perturbation operators for various solution encodings. The gene depicted in the blue color is mutated.}
    \label{fig:encodings}
\end{figure}

The pseudocode for evolutionary algorithms is given in Algorithm~\ref{alg:ea}, while in Figures~\ref{fig:float_encoding} and~\ref{fig:binary_encoding}, we present mutations working on floating-point and binary encoding, respectively.

\begin{algorithm}
\small
\caption{Pseudocode for EA.}
\label{alg:ea}
\begin{algorithmic}
\STATE $t\leftarrow 0$
\STATE $P(0) \leftarrow CreateInitialPopulation$
\REPEAT
	\STATE $t \leftarrow t + 1$
	\STATE $P'(t)\leftarrow SelectionMechanism\ (P(t - 1))$
	\STATE $P(t)\leftarrow VariationOperators (P'(t))$
\UNTIL{\textit{TerminationCriterion}}
\STATE $Return \ Optimal Solution Set (P)$
\end{algorithmic}
\end{algorithm}

\subsection{Experiments on Encoding \& RNG}
\label{app:preliminary_experiments}
We depict the evaluation results for different encodings and RNGs in~\Cref{tab:eval_speed_appendix}
\begin{table}[h]
    \centering
    \caption{Averaged executions per second reached with the respective variation of \coolname. Positive percentages that \coolname was this much faster than the fuzzer in the column.}
    \footnotesize
    \begin{tabular}{|l|r|r|r|}
    \hline
        & \multicolumn{1}{|c}{\coolname}& \multicolumn{1}{|c}{D-Std. RNG} & \multicolumn{1}{|c|}{D-Real Valued} \\
    \hline
     Benchmark   &   execs/s &  execs/s &   execs/s \\
    \hline
    \texttt{cxxfilt}      & 2210.41 & 2151.47 & 1860.02 \\
    \texttt{objcopy}     & 2610.73  & 2630.80 & 2678.77 \\
    \texttt{objdump}     & 1687.52  & 2161.75 &   2225.95  \\
    \texttt{readelf}     & 3405.55  & 2711.17 &   2815.48      \\
    \texttt{size}        & 3140.08  & 2733.33 &   2910.92   \\
    \texttt{strip}       & 2686.19  & 2492.25 & 2665.44   \\\hline
     geomean     &             & +3.62 \% &  +2.42\%     \\
    
    \hline
    \end{tabular}

    \label{tab:eval_speed_appendix}
\end{table}

\subsection{Seed Used for Binutils}
\label{app:binutils_seed}
We build a minimal ELF seed testcase for binutils to achieve adequate execution speed. Its code is depicted in~\Cref{lst:binutils_seed}.
\begin{lstlisting}[label=lst:binutils_seed,caption={Source code for binutils seed, calling \texttt{sys\_exit}.},captionpos=b,style=CStyle]
extern "C" void _start() {
     __asm("mov $60, %
            xor %
            syscall");
}
\end{lstlisting}

\subsection{LAVA-M - Finding Known Bugs}
\label{subsec:eval_lava}

LAVA-M~\cite{dolan2016lava} is a synthetic set of bugs inserted into the GNU coreutils suite. These hard-to-reach bugs are injected automatically into the real-world binaries \texttt{who}, \texttt{uniq}, \texttt{md5sum}, and \texttt{base64}. While LAVA-M has questionable implications on real-world performance, it is commonly used to evaluate fuzzers in research~\cite{redqueen,vuzzer,tfuzz,angora,neufuzz,mopt}.
As LAVA-M is heavily focusing on comparisons, LAVA-M favors approaches that concentrate on improving mutation operators themselves~\cite{redqueen}. 
Hence, we keep this for the sake of completeness here in the abstract.
While the benchmark provides one initial test case per target, we added an uninformed, empty test case for each target to be consistent with our other experiments.
Each target is fuzzed for five hours, as commonly done for the LAVA-M benchmark in fuzzing papers~\cite{mopt,redqueen,vuzzer}.
\Cref{tab:lava_crashes} depicts the results for \coolname, MO\small{PT}\normalsize, and AFL over three runs.
Notably, \coolname is the only fuzzer in our evaluation that finds bugs across all targets and consistently finds the highest number of bugs in each target.
For \texttt{uniq} and \texttt{who}, which are the only targets where all fuzzers found bugs, we further analyze in which fuzzing loop stage the bugs were found.
In the case of \texttt{uniq}, \coolname finds 50\% of the bugs using the havoc stage, while MO\small{PT} \normalsize and AFL exclusively found all bugs using splicing.
For \texttt{who}, the havoc stage attributes for one-third of the bugs found by \coolname, whereas on AFL, the havoc stage accounts for 20\% of the bugs. On MO\small{PT}\normalsize, the havoc stage is never successful in finding a bug, possibly because some mutators are never scheduled.
By comparing the maximum numbers found per fuzzer, we can conclude that \coolname found more bugs than just the overlap between all fuzzers.
Finally, \coolname's approach for mutation scheduling is orthogonal to, e.g., improvements in overcoming branch checks~\cite{angora,qsym,redqueen}, and can be used to optimize the scheduling of the respective mutation operators to achieve a synergetic effect.

\begin{table}[h]
    \centering
    \caption{Crashes found in LAVA-M, average crashes over three runs as well as the highest number of crashes encountered within an individual run.}
    \footnotesize
    \begin{tabular}{|l|r|r|r|r|r|r|}
    \hline
                 & \multicolumn{2}{|c}{\coolname}& \multicolumn{2}{|c}{MOPT}& \multicolumn{2}{|c|}{AFL}\\
    \hline
    Benchmark   &   Avg. &   Max. &   Avg. &   Max. &   Avg. &   Max.\\
    \hline
    \texttt{base64} & 1 & 2 & 0 & 0 & 0.33 & 1 \\
    \texttt{md5sum} & 0.33 & 1 & 0.33 & 0 & 0 & 0 \\
    \texttt{uniq}   & 3.67 & 4 & 0.33 & 1 & 0.33 & 1 \\
    \texttt{who}    & 3 & 3 & 2 & 2 & 2.67 & 3 \\
     \hline
    Total & 8 & 10 & 2.67 & 3 & 2.33 & 1 \\
    \hline
    \end{tabular}
    \label{tab:lava_crashes}
\end{table}

\subsection{Mutations in the AFL Havoc Stage}
\label{app:afl_mutations}

Table~\ref{tab:mut_operators} lists all mutations defined in the AFL havoc stage.

\begin{table}[h]
  \centering
  \small
    \begin{tabularx}{\linewidth}{rX}
    \multicolumn{1}{l}{\textbf{ID}} & \textbf{Description} \\
    \toprule
    0     & Flip single bit \\
    1     & Set byte to interesting value \\
    2     & Set word to interesting value \\
    3     & Set dword to interesting value \\
    4     & Randomly subtract from byte \\
    5     & Randomly add to byte \\
    6     & Randomly subtract from word \\
    7     & Randomly add to word \\
    8     & Randomly subtract from dword \\
    9     & Randomly add to dword \\
    10    & Set a random byte to a random value \\
    11    & Delete Bytes \\
    12    & Delete Bytes \\
    13    & Clone bytes (75\%) or insert a block of constant bytes (25\%) \\
    14    & Overwrite bytes with a randomly selected chunk (75\%) or fixed bytes (25\%) \\
    15    & Overwrite bytes with an extra \\
    16    & Insert an extra \\\bottomrule
    \end{tabularx}%
  \vspace{2pt}
  \caption{Mutations defined in the AFL havoc stage, descriptions taken from the AFL source code~\cite{afl}. Extra refers to target-specific dictionary entries. 11 and 12 trigger the same mutation to increase selection probability based on practical experience.}
  \label{tab:mut_operators}%
\end{table}%

\end{document}